\documentclass[aps,prb,twocolumn,floatfix,showpacs,superscriptaddress,fleqn]{revtex4-1}
\usepackage{amssymb}

\usepackage{graphicx}
\usepackage{amsmath}
\usepackage{bm}
\usepackage{color}
\usepackage{braket}
\usepackage{array}
\usepackage[breaklinks]{hyperref}
\usepackage[all]{hypcap}

\hypersetup{
    plainpages=false,
    bookmarks=false,         
    unicode=false,          
    pdfborder={0 0 0}
    pdftoolbar=true,        
    pdfmenubar=true,        
    pdffitwindow=false,     
    pdfstartview={FitH},    
    pdftitle={PRB},    
    pdfauthor={Vache Folle},     
    pdfproducer={LNCMI-CNRS-UGA-UPS-INSA}, 
    pdfkeywords={High} {Magnetic} {Fields}, 
    pdfnewwindow=true,      
    linktoc=section,
    colorlinks=true,       
    linkcolor=blue,          
    citecolor=red,        
    filecolor=magenta,      
    urlcolor=blue           
}

\begin{document}

\author{A. Poux}
\affiliation{Laboratoire National des Champs Magn\'etiques Intenses, CNRS-UGA-UPS-INSA, 143 avenue de Rangueil, 31400
Toulouse, France}

\author {Z. R. Wasilewski}
\affiliation{Institute for Microstructural Sciences, National Research Council, Ottawa, Canada} \affiliation{Department
of Electrical and Computer Engineering, University of Waterloo, Canada.}\affiliation{Department of Physics and
Astronomy, Waterloo Institute for Nanotechnology, Waterloo, Cananda}

\author{K. J. Friedland}
\affiliation{Paul Drude Institut f\"ur Festk\"orperelektronik, Berlin, Germany}

\author{R. Hey}
\affiliation{Paul Drude Institut f\"ur Festk\"orperelektronik, Berlin, Germany}

\author{K. H. Ploog}
\affiliation{Paul Drude Institut f\"ur Festk\"orperelektronik, Berlin, Germany}

\author{R. Airey}
\affiliation{Department of Electronic and Electrical Engineering, University of Sheffield, UK}

\author{P.\ Plochocka}
\affiliation{Laboratoire National des Champs Magn\'etiques Intenses, CNRS-UGA-UPS-INSA, 143 avenue de Rangueil, 31400
Toulouse, France}

\author{D. K. Maude}\email{duncan.maude@lncmi.cnrs.fr}
\affiliation{Laboratoire National des Champs Magn\'etiques Intenses, CNRS-UGA-UPS-INSA, 143 avenue de Rangueil, 31400
Toulouse, France}

\title{A microscopic model for the magnetic field driven breakdown of the dissipationless state in the integer and fractional quantum Hall effect}


\date{\today}

\begin{abstract}
Intra Landau level thermal activation, from localized states in the tail, to delocalized states above the mobility edge
in the same Landau level, explains the $B_c(T)$ (half width of the dissipationless state) phase diagram for a number of
different quantum Hall samples with widely ranging carrier density, mobility and disorder. Good agreement is achieved
over $2-3$ orders of magnitude in temperature and magnetic field for a wide range of filling factors. The Landau level
width is found to be independent of magnetic field. The mobility edge moves, in the case of changing Landau level
overlap to maintain a sample dependent critical density of states at that energy. An analysis of filling factor
$\nu=2/3$ shows that the composite Fermion Landau levels have exactly the same width as their electron counterparts. An
important ingredient of the model is the Lorentzian broadening with long tails which provide localized states deep in
the gap which are essential in order to reproduce the robust high temperature $B_c(T)$ phase observed in experiment.
\end{abstract}

\maketitle

\section{Introduction}

The integer quantum Hall effect\cite{Klitzing1980,Klitzing1986} (QHE) can in principle be understood within the
framework of a single particle picture. The quantized plateaux in the Hall resistance $R_{xy}$, together with the zero
resistance state in the longitudinal resistance $R_{xx}$ occur whenever the Fermi energy lies in a gap in the density
of states. When a two dimensional electron gas (2DEG) is placed in a perpendicular magnetic field, the movement of the
Fermi energy through the quantized Landau levels is driven by the $eB/h$ degeneracy of a spin Landau level. For a
two-dimensional carrier density $n_s$, the filling factor is defined as $\nu=n_s h/eB$. At even filling factors $E_f$
lies in a cyclotron gap while for odd filling factors it lies in a spin gap. This picture only gives rise to quantum
Hall states of non zero width in the presence of disorder broadened Landau levels, with localized states in the tails,
beyond a sharp mobility edge. As we move away from exact integer filling factor, the dissipationless resistance is
quenched once the increasing (decreasing) $eB/h$ degeneracy drives the Fermi energy into the delocalized states near
the Landau level center.

Thus, understanding the nature of the disorder is important if we are to fully understand the quantum Hall effect, with
important implications for metrology. Disorder acts subtly, it simultaneously strengthens the integer QHE (wider
plateau, larger critical current), while competing with or even annihilating the fractional QHE. For example, the
competition between the disorder induced energy cost of reversing spins, and the exchange energy gain, controls the
opening of the many body spin gap at odd filling factors.\cite{Leadley1998,Piot2005} Indeed, the shape of the disorder
broadened, Landau levels remains controversial. In an exquisitely difficult experiment, a detailed analysis of the saw
tooth form of the extremely small oscillations in the 2DEG magnetization versus magnetic field, essentially the weight
of the higher harmonic terms in a Liftshitz-Kosevich approach, suggested that the Landau level broadening is
Lorentzian.\cite{Potts1996} However, subsequent work led to diverging conclusions.\cite{Zhu2003,Usher2009} The presence
of disorder is also naturally required to explain the fractional
QHE\cite{Tsui1982,Laughlin1983,Willett1987,Stormer1999,Stormer1999a} within the composite Fermion framework, in which
the fractional filling factors for electrons map to integer filling factors on non interacting composite Fermions,
quantized into Landau levels by an effective quantizing magnetic field $B^*=B - B_{1/2}$.\cite{Jain1989}

\begin{figure}[]
\begin{center}
\includegraphics[width= 8.5cm]{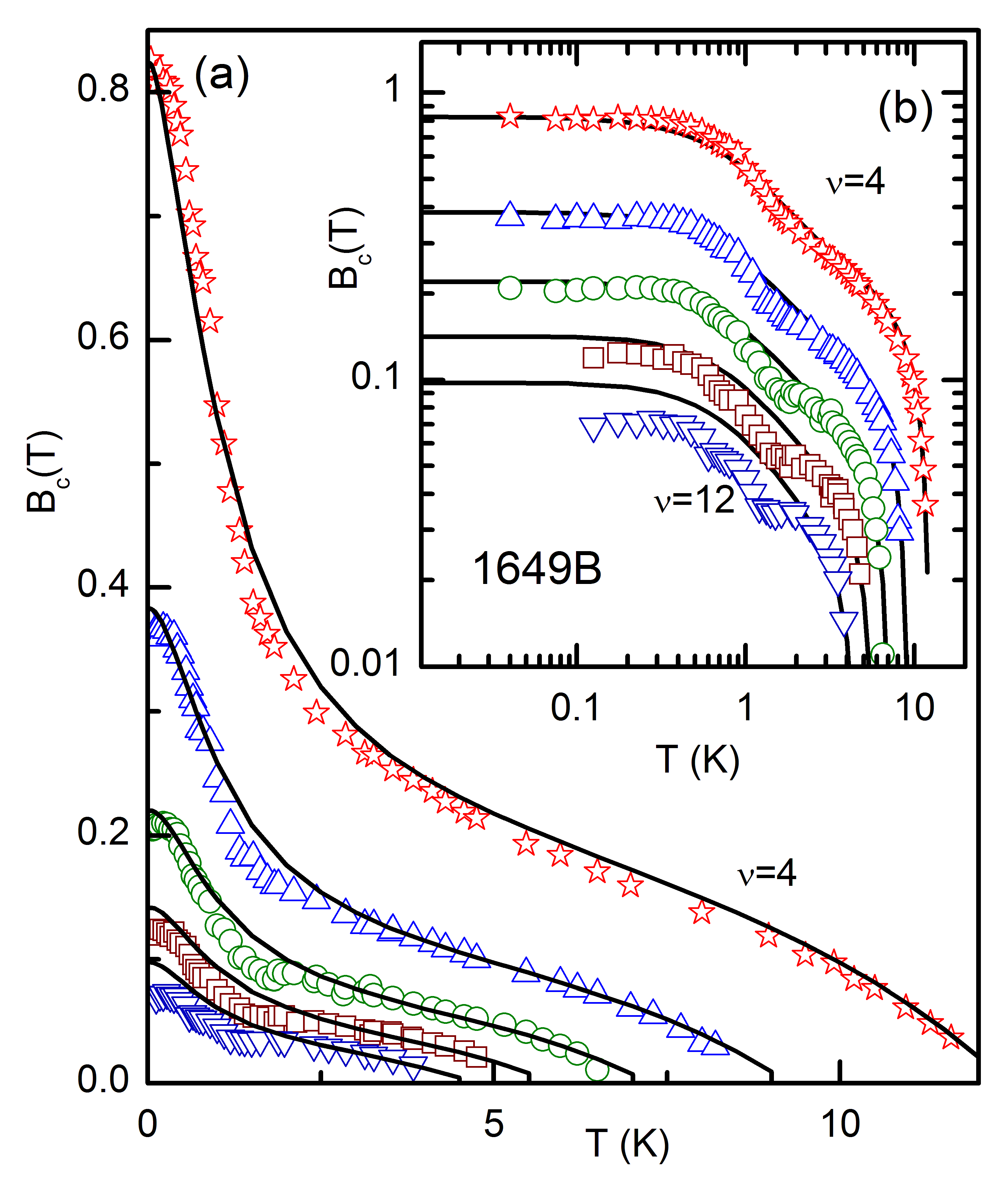}
\end{center}
\caption{(color online) (a) Critical magnetic field (half width of the dissipationless region) versus temperature for
integer quantum Hall filling factors $\nu=4,6,8,10,12$ for sample 1649B. The data is taken from
reference~[\onlinecite{Rigal1999}]. The solid curves are the calculated critical magnetic field due to thermal
activation of carriers to delocalized states above the mobility edge in the disorder broadened Landau level. (b) A log
log plot of the same data and calculations. }\label{fig:Even1649Barry}
 \end{figure}

It has been shown that the width in magnetic field ($\Delta B$) of the maxima in $R_{xx}$, the so called plateau to
plateau transitions in $R_{xy}$, follows a universal scaling law with $\Delta B \propto T^{\kappa}$ where $\kappa=0.42$
is universal.\cite{Wei1988,Pruisken1988} The scaling law was derived using renormalization group theory (RGT), which is
generally applied to problems (phase transitions) which are to complicated to solve from first principles. The
universality is the expected signature of a quantum phase transition.\cite{Sondhi1997} However, the experimental
universal value of $\kappa=0.42$ turned out to be controversial, with some later work establishing a universal value of
$\kappa=0.58$ which was tentatively attributed a non Fermi liquid like
behavior.\cite{McEuen1990,Koch1991,Lee1993,Coleridge1994,Lee1994,Pan1997,Shahar1997,Shahar1998,Coleridge1999,Ponomarenko2004,Ponomarenko2005,Lang2005,Visser2006,Lang2007}

In a different approach, Rigal and coworkers reported that the phase diagram for the (half) width of the
dissipationless state, in the longitudinal resistance $R_{xx}$, versus magnetic field of a quantum Hall sample bears a
remarkable resemblance to the phase diagram for the critical magnetic field of a high temperature superconductor with
vortex melting.\cite{Rigal1999} This can be seen in figure\,\ref{fig:Even1649Barry} where we replot the $B_c(T)$ phase
diagram with the data taken from Ref.[\onlinecite{Rigal1999}]. As discussed by Rigal \emph{et al.}, all even integer
filling factors show a low temperature Gorter-Casimir like phase with, perhaps surprisingly, the same (extrapolated)
critical temperature $T_c^{LT} \simeq 1.5K$. In the high temperature phase the critical temperature $T_c^{HT}$ for even
filling factors scales approximately as the cyclotron gap. The low temperature values of the critical magnetic field
scale as $B_c(\nu) = B_{c0}(1/\nu^2 - 1/\nu_0^2)$ consistent with Landau levels with a constant (filling factor
independent) ratio of the number of localized to delocalized states within a Landau level. Here $\nu_0$ is the filling
factor above which the conduction is no longer dissipationless at $T=0$\,K. The $1/\nu^2$ scaling arises from the
Landau level degeneracy $eB_f/h\nu$ (magnetic field $B=B_f/\nu$ where $B_f$ is the magnetic field at which $\nu=1$
occurs), and the fact that for every magnetic flux quanta added (or removed), $\nu$ electrons disappear (or appear) in
the Landau level at the Fermi energy (since there are $\nu$ occupied Landau levels below or at $E_f$ which gain or lose
an electron).

In this paper, we show that the $B_c(T)$ phase diagram of Rigal \emph{et al.} can be exactly explained by a simple
model involving thermal activation from localized states in the tails, to delocalized state near the center, of a
disorder broadened Landau level at the Fermi energy. This work is then extended, to a number of different 2DEG samples
with very different carrier densities and mobilities, for both even, odd and fractional filling factors. Our results
strongly suggests that Landau level broadening is Lorentzian. The long Lorentzian tails are a fundamental requirement
to explain the robust high temperature $B_c(T)$ phase. In other words, it is essential to have a significant number of
states, deep in the gap, into which the Fermi level can be pushed in order to suppress thermal activation at the high
critical temperatures observed in experiment. We will see that Gaussian broadening simply fails to meet this stringent
requirement.

We stress that the Lorentzian line shape is found for all the samples investigated, including a high mobility - low
disorder 2DEG which shows well developed fractional quantum Hall states. The model can equally well explain the
$B_c(T)$ phase diagram for odd and fractional filling factors, with a self consistent limited parameters set, for each
sample, explaining the data at all filling factors. From the data set and model, a global and coherent picture emerges,
in which the mobility edge, normally assumed to be fixed, can move under conditions of changing Landau level overlap.
This occurs for example, in case of (i) the emergence of a single particle spin gap in the Landau level at the Fermi
energy for even filling factors, (ii) the emergence of a many body spin gap at odd filling factors and (iii) in the
case of the overlap of localized states in the tails of Landau levels at low magnetic fields. This can all be
understood within the framework of conduction via variable range
hopping\cite{,Mott1969,Efros1975,Ono1982,Briggs1983,Polyakov1993a,Polyakov1993,Hohls2002} in the Landau level tails.
Overlapping the \emph{localized} states of different Landau levels increases the density of states at the Fermi energy,
increasing the probability of variable range hopping \emph{i.e.} causes the mobility edge to shift further away from
the center of the Landau level. Thus, what is required to have a robust quantum Hall sample is large disorder, in order
to have a large Landau level broadening, with an even larger gap to avoid at all cost the overlap of the localized
tails. This probably goes a long way to explain the current interest in graphene for
metrology.\cite{Tzalenchuk2010,Janssen2013,Janssen2015}

The rest of the paper is arranged as follows: In section\,\ref{Sec:TherAct} the intra Landau level thermal activation
model using Fermi-Dirac statistics is presented and demonstrated to work well for the previously published $B_c(T)$
phase diagram in Ref.\,[\onlinecite{Rigal1999}]. This result is then extended to other samples. In
section\,\ref{Sec:ExptTech} a brief description of the experimental techniques and sample characteristics are
presented. New results, for three very different 2DEG samples, are presented in section\,\ref{Sec:ExptResults}. The
thermal activation model is shown to work for even, odd and fractional filling factors in widely different samples. For
certain samples, in which the Landau level overlap is changing, a clear signature of the movement of the mobility edge
is observed. In section\,\ref{Sec:ScalingTheory} the implications of this work for scaling theory are discussed.
Finally, global conclusions are drawn and the implications and outlook for future work are drawn in
section\,\ref{Sec:Conclusion}.

\section{Intra Landau level thermal activation to the mobility edge}\label{Sec:TherAct}

\begin{figure}[]
\begin{center}
\includegraphics[width= 7.5cm]{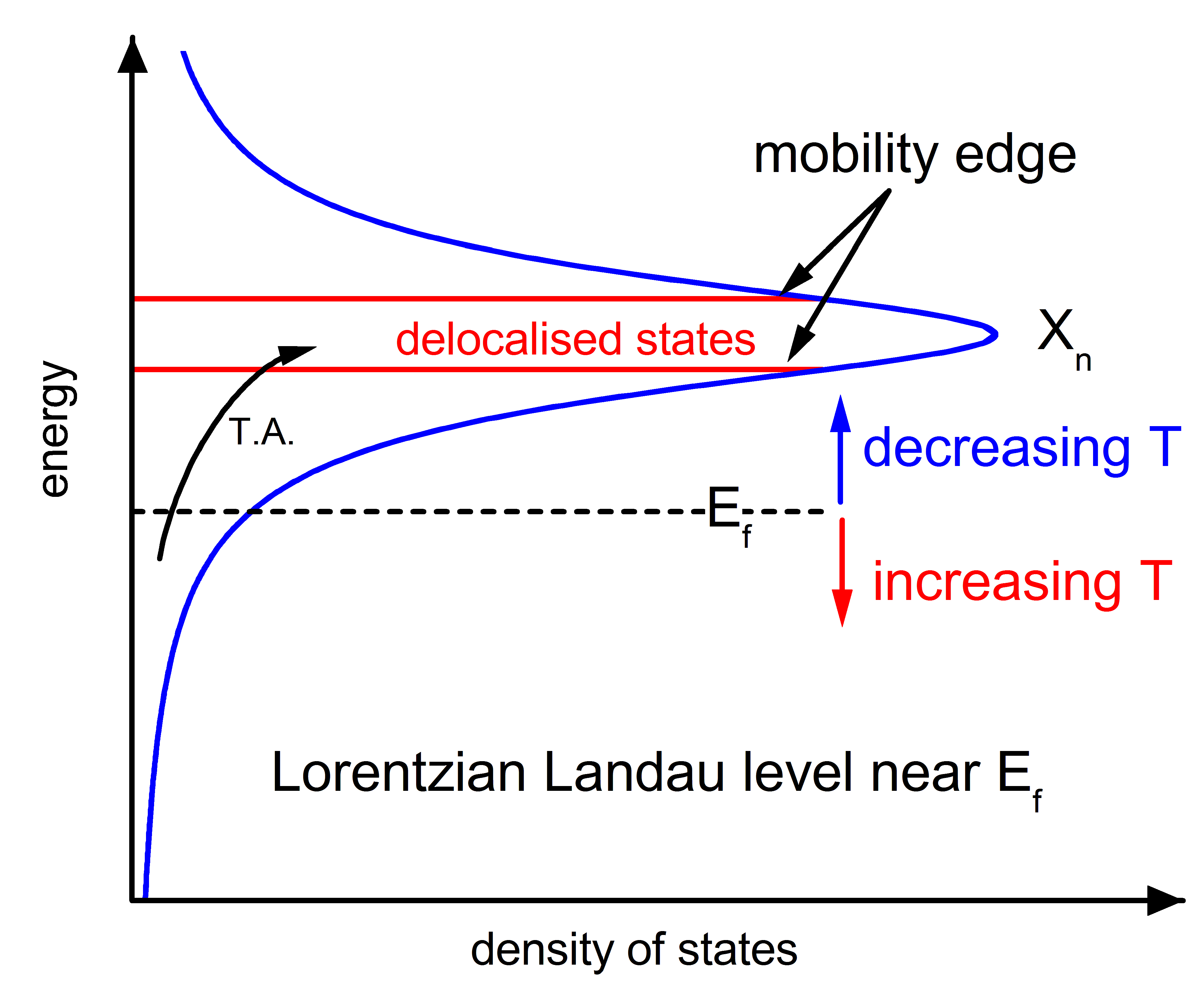}
\end{center}
\caption{(color online) Schematic showing intra Landau level thermal activation with a Lorentzian broadening. As
sketched the single particle Zeeman energy is much smaller than the broadening so that spin splitting can be neglected.
The required movement of $E_f$ with temperature, in order to maintain the dissipationless state, is indicated by the
vertical arrows. In the experiment the Fermi energy is moved by changing the magnetic field (filling factor).
}\label{fig:Lorentzian}
 \end{figure}

We propose a simple microscopic model, based on thermal activation \emph{within} the partially occupied Landau level at
the Fermi energy, which can explain both the low and high temperature phases in the $B_c$ versus $T$ phase diagram for
\emph{all filling factors}. A Lorentzian broadening of the Landau level is used, with the assumption that states in the
tails are localized and do not contribute to transport, while states in the center, above a sharp mobility edge are
delocalized. For a given integer filling factor, at the critical temperature the Fermi level lies in the center of the
gap and the zero resistance state is destroyed due to a thermally activated (critical) population $\nu_c$ of
delocalized states just above the mobility edge. This explains why the high temperature phase has a critical
temperature $T_c^{HT}(\nu)$ which scales approximately with the cyclotron gap for even filling factors. With decreasing
temperature, it is possible to push the Fermi energy up, closer to the mobility edge, before the critical occupation
$\nu_c$ is obtained (see Fig.\,\ref{fig:Lorentzian}). At $T=0$, to a good approximation, the Fermi energy coincides
with the mobility edge. Thus, the low temperature value of $B_c$ is simply controlled by the fraction of delocalized
states in the Landau level. An important ingredient of this model is the line shape used to describe the disorder
broadening of the Landau level. The robustness of the $B_c$ versus $T$ phase diagram in
reference\,[\onlinecite{Rigal1999}] results from the Lorentzian Landau level broadening, with its extremely long tails
which provide a wide range of possible Fermi energies, and hence a wide range of temperature over which the
dissipationless regime is maintained. Other line shapes (\emph{e.g.} Gaussian) have short tails and are unable to
reproduce the robust high temperature phase. An important part of this work is extending this result to a range of
samples with very different mobilities and disorder.

The spin Landau levels are described using two Lorentzians,
\begin{equation}
g_{n\uparrow\downarrow}(E) = \frac{1}{\pi \Gamma \left(1 +
\left(\frac{E-X_{n\uparrow\downarrow}}{\Gamma}\right)^2\right)},\nonumber\\
\end{equation}
where,
\begin{equation}
 X_{n\uparrow\downarrow} = \left(n + \frac{1}{2}\right)\hbar \omega_c \pm \frac{1}{2} g^* \mu_B B\nonumber.
\end{equation}
Each Lorentzian is centered at $X_{n\uparrow\downarrow}$ with a full width at half maximum of $2 \Gamma$. The
Lorentzians, as written, are normalized so that the integral over all energies $\int g_{n\uparrow\downarrow}(E)dE = 1$
and all calculation are performed using filling factor rather than carrier density.

For a given position of the Fermi energy $E_f$, the filling factor can be obtained using
\begin{equation}\label{Eq:nu_tot}
\nu(T,E_f) = \int_{-\infty}^{\infty} \sum_{n=0}^\infty [g_{n\uparrow}(E) + g_{n\downarrow}(E)]f(E)dE,
\end{equation}
where $f(E)=1/(1 + \exp((E-E_f)/KT))$ is the Fermi-Dirac distribution function. The integral is computed numerically
and to speed up the calculation we consider only Landau levels in the direct vicinity of the Fermi energy (\emph{i.e.}
the Landau levels immediately above and below $E_f$). Other Landau levels are assumed to be either full or empty.
Knowing the filling factor we can calculate the magnetic field which corresponds to the current position of the Fermi
energy, $B=B_f/\nu$ where $B_f=n_s h/e$ is the magnetic field at which filling factor $\nu=1$ occurs in a sample with
carrier density $n_s$.

In this simple model, the dissipationless resistance state ceases to exist when the thermally activated population of
delocalized states above the mobility edge exceeds a critical value (which depends on the value of the critical
resistance $R_c$ used to determine the half width ($B_c$) of the dissipationless state). At low temperatures, this
thermal activation occurs from electrons occupying states below the mobility edge in the \emph{same Landau level}. This
explains why the low temperature phase has the same critical temperature ($T_c^{LT}\simeq1.5K$) for all filling
factors, \emph{i.e.} it does not depend on the size of the cyclotron gap. With increasing temperature, the Fermi energy
is pushed further and further down into the tail of the Landau level, in order to maintain the occupation of the
delocalized states below the critical value. Thus, the width in temperature, of the low temperature phase is very
sensitive to the size ($\Gamma$) of the Landau level broadening. At $T=0$, the resistance lifts off, when the mobility
edge coincides with the Fermi energy, so that $B_c(T=0)$ depends only on the number of localized states below the
mobility edge \emph{i.e.} $\Gamma/\Gamma_{dl}$.

Thermal broadening is included phenomenologically by writing
\begin{eqnarray}
\Gamma(T) &=& \sqrt{\Gamma(0)^2 + (\alpha KT)^2},\nonumber\\
\Gamma_{dl}(T) &=& \Gamma_{dl}(0) \sqrt{1 + (\alpha KT/\Gamma(0))^2},\nonumber
\end{eqnarray}
where $\alpha \sim 1$ is a dimensionless parameter. Note that the ratio $\Gamma_{dl}/\Gamma$ is independent of the
temperature.

For a given even integer filling factor ($\nu=2n$), temperature and position of $E_f$, the population of delocalized
states above the mobility edge in the Landau level directly above the Fermi energy is given by,

\begin{eqnarray}\label{Eq:nu_c}
\delta\nu (T, E_f) = \int_{X_{n\uparrow}-\Gamma_{dl}}^{X_{n\uparrow}+\Gamma_{dl}} g_{n\uparrow}(E) f(E) &dE&\nonumber\\
+ \int_{X_{n\downarrow}-\Gamma_{dl}}^{X_{n\downarrow+\Gamma_{dl}}} g_{n\downarrow}(E) f(E) &dE&.
\end{eqnarray}

The critical temperature in the high temperature phase, which scales as the cyclotron energy, corresponds to the
situation where $E_f$ lies exactly at midgap and the resistance starts to lift off due to thermal activation across the
gap. While the critical population of delocalized states could be taken as a fitting parameter, it makes more sense to
use experiment to fix this value; for each filling factor we use the observed critical temperature of the high
temperature phase, to calculate the (critical) population, $\delta\nu_c$, of delocalized state above the mobility edge,
when $E_f$ lies in the center of the cyclotron gap.

Assuming that the obtained value of $\delta \nu_c$, for a given filling factor, does not depend on the temperature, we
calculate iteratively for each temperature, the required position of $E_f$ using Eq.\ref{Eq:nu_c} to have the critical
occupation $\delta \nu_c$ of the delocalized states. A valid solution is found if $E_f$ lies somewhere between midgap
and the center of the Landau level. Then, using the value of $E_f$ we calculate using Eq.\ref{Eq:nu_tot} the filling
factor and hence the magnetic field. We calculate the cyclotron energy $\hbar\omega_c=\hbar e B/m^*$ and the Zeeman
energy $g \mu_B B$ using the magnetic field $B=B_f/\nu$ corresponding to the even integer filling factor. Within this
approximation, the problem is symmetric with respect to moving away from integer filling factor to lower or higher
fields \emph{i.e.} the partial filling factor, $\varepsilon$, (when $\delta \nu = \delta \nu_c$) of electrons in the
previously empty, of holes in the previously full Landau level is the same. In other words the filling factors found
are $\nu + \varepsilon$ and $\nu - \varepsilon$ corresponding to magnetic fields $B_1=B_f/(\nu + \varepsilon)$ and
$B_2=B_f/(\nu -\varepsilon)$ with $B_c = (B_2 - B_1)/2$. A second iteration, using this time the previous values of
$B_1$ and $B_2$ to calculate the cyclotron and Zeeman energies only changes $B_c$ by a few mT and was thus judged to be
unnecessary. The approximation works well because (i) $B_1$ and $B_2$ are generally not too different from $B_f/\nu$
($\nu$ is an even integer) and (ii) the corrections to $B_1$ and $B_2$ actually tend to cancel. The cyclotron energy is
calculated using the accepted effective mass in GaAs, corrected for non parabolicity in the higher density
samples.\cite{Raymond1979} To calculate the Zeeman energy we use the accepted value of the Land\'e g-factor in bulk
GaAs $g^*=-0.44$ which is different from 2 due to spin orbit coupling.\cite{Weisbuch1977}

In the model we have three fitting parameters; (i) the Landau level broadening $\Gamma$ which is adjusted to have the
correct width of the low temperature phase, (ii) the position of the mobility edge (number of localized states) which
is controlled by $\Gamma_{dl}$, essentially the ratio $\Gamma_{dl}/\Gamma$ determines the value of $B_c(T \rightarrow
0)$, and (iii) the thermal broadening parameter $\alpha$ which improves the agreement at high temperatures. The
experimental value of $T_c^{HT}$ used to define $\delta \nu_c$ ensures that $B_c \rightarrow 0$ at the correct
temperature. The solid lines in figure\,\ref{fig:Even1649Barry} are calculated using $\Gamma=3.0$\,K, $\Gamma_{dl}=0.9
\Gamma$, $\alpha=1$, $m^*=0.072 m_e$ and $g^*=-0.44$. As can be seen in the log log plot in
figure\,\ref{fig:Even1649Barry}(b), this single parameter set provides a good fit to all even filling factors, over
nearly three order of magnitude in temperature and almost two orders of magnitude in magnetic field.

\begin{figure}[]
\begin{center}
\includegraphics[width= 8.5cm]{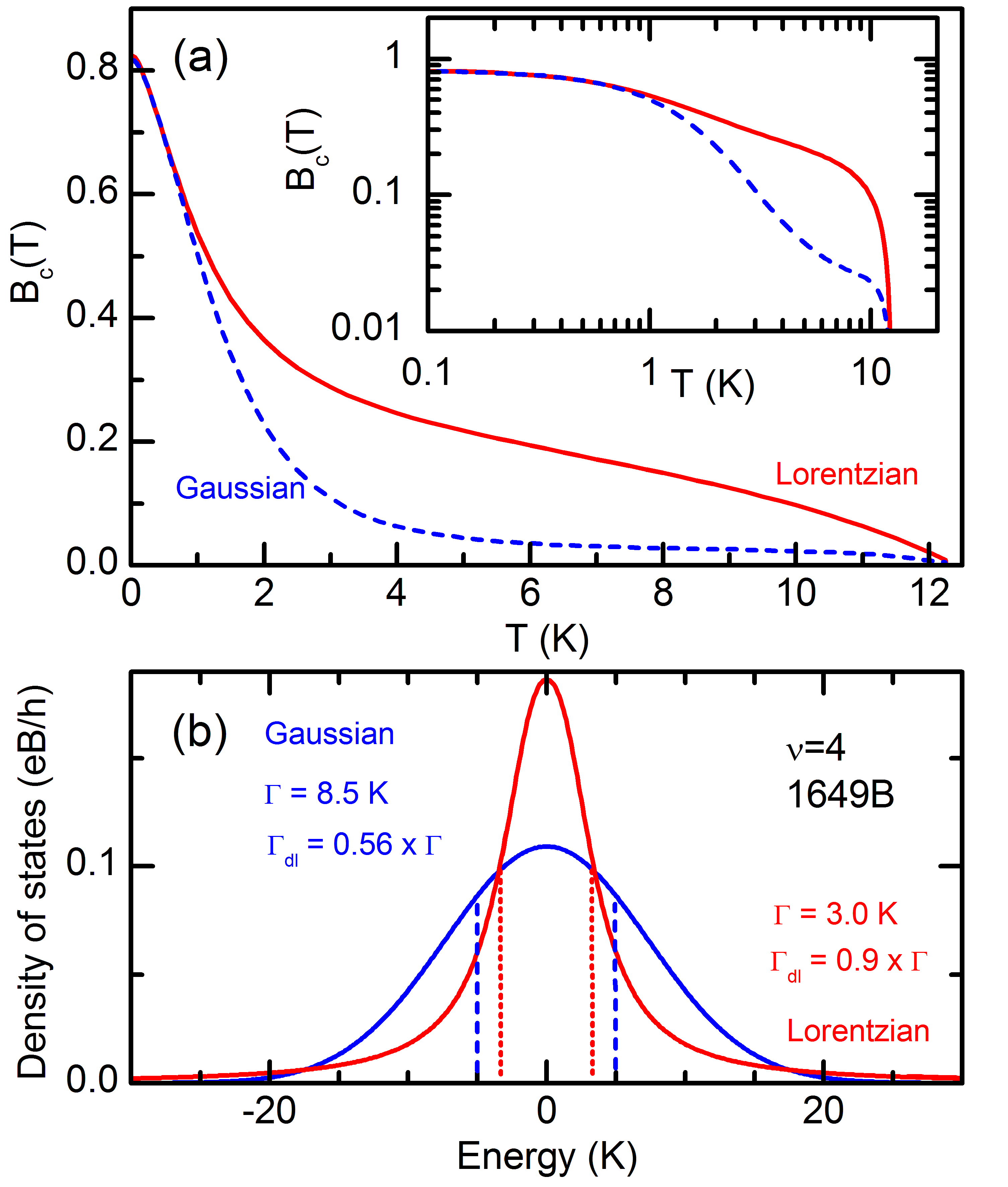}
\end{center}
\caption{(color online) (a) Calculated $B_c(T)$ for Lorentzian and Gaussian broadened Landau levels. For the Lorentzian
we have used the parameters and conditions of $\nu=4$ in sample 1649B. The width of the Gaussian is precisely
determined by the width of the $B_c(T)$ for $T<1$\,K and completely fails to reproduce the data for $T>1$\,K. The inset
shows a log log plot of the same curves. (b) The Lorentzian and Gaussian Landau levels drawn according to the
parameters used in the fits. Vertical broken lines indicates the approximate position of the mobility edge. Note spin
splitting is not resolved for either line shape.}\label{fig:Laurrazian}
 \end{figure}

Similar calculations using a Gaussian line shape reveal the importance of the long Lorentzian tails for the robustness
of the high temperature part of the $B_c(T)$ phase diagram. As an example in figure\,\ref{fig:Laurrazian}(a) we show
calculations for a Lorentzian broadening performed with the same conditions and parameter set as for $\nu=4$ for sample
1649B (which fit the data perfectly). For calculations using the Gaussian broadened Landau level, the width of the
Landau level is precisely determined by the width of the low temperature part of the $B_c(T)$ phase diagram where $B_c$
is rapidly falling, and the width of the delocalized states (position of the mobility edge $\Gamma_{dl}$) is determined
by the value of $B_c(T \rightarrow 0)$. For $T>1$\,K the predicted value of $B_c$ are significantly lower than required
and for $T>4$\,K the predicted $B_c$ falls below the experimental limit at which we can measure, \emph{i.e.} the high
temperature phase is predicted to be washed out. The short Gaussian tails do not have sufficient states with energies
well below the mobility edge in order that $B_c$ survives up to the temperature at which activation across the
cyclotron gap finally kills the dissipationless state. The Gaussian and Lorentzian Landau levels are sketched in
figure\,\ref{fig:Laurrazian}(b) with the broadening used in the calculations. The position of the mobility edge are
indicated by the vertical broken lines. Note that the parameters given for the Landau level broadening ($\Gamma$) is
half of the full width at half maximum (FWHM).

For the moment we have focussed uniquely on the simplest case of even integer filling factors, when the Fermi energy
lies in the cyclotron gap, the system is spin unpolarized, so that the spin gap is given by the single particle Zeeman
energy $g^* \mu_B B$ and many body effects can safely be neglected. However, it is trivial to extend the thermal
activation model to odd filling factors, where the Fermi energy lies in the spin gap. This gap can be reproduced using
an effective g-factor $g^*_{eff}>>g^*$ due to the many body enhancement of the spin gap in the spirit of the Ando
model.\cite{Ando1974} Equally, the model can be extended to fractional states, within the composite fermion
framework.\cite{Jain1989} A crucial test of the thermal activation model will be its ability to explain, the $B_c(T)$
phase diagram for even, odd and fractional filling factors in different samples with widely different characteristics.
Such an investigation will be presented in the the following sections.

\section{Experimental techniques}\label{Sec:ExptTech}

For the electrical transport measurements the sample was placed in the mixing chamber of a top loading dilution
refrigerator equipped with a $16$\,T superconducting magnet. The sample and wires are immersed directly in the
He$^3$/He$^4$ mixture. During the top loading, the sample is cooled slowly over a number of hours. In contrast to
previous measurements all data was taken during the same cool down in the dilution refrigerator. Temperatures in the
range $10$\,mK to $4.2$\,K were obtained with the sample in liquid. For higher temperatures up to $\simeq12$\,K the
mixture was removed with the exception of a few mbars of exchange gas. The temperature was controlled either using a
heater mounted directly in the mixing chamber a few cm from the sample, and/or by controlling the amount of He exchange
gas in the inner vacuum chamber of the refrigerator. This was achieved via the temperature of a sorb pump (small piece
of activated charcoal) placed in the vacuum chamber.

Hall bars where fabricated using standard photolithography techniques with and aspect ratio of $3:1$ $e.g.$ a Hall bar
width of $250 \mu$m with $750 \mu$m between the voltage contacts. A constant low current of $10-20$\,nA at 10.66\,Hz
was applied using a 100M$\Omega$ series resistor and the oscillator output of an SR830 lockin amplifier. Low pass
preamplifiers, based on the INA111 Burr Brown low noise operational amplifier, where placed as close to the sample
probe as feasible to keep cable lengths below 50\,cm for the unamplified signal. The longitudinal resistance $R_{xx}$
was measured using phase sensitive detection. The sample temperature was monitored using Ru0$_2$ ($T<4.2$\,K) and
Cernox ($T>4.2$\,K) thermometers placed close to the sample. The thermometers were also measured using preamplifiers
and phase sensitive detection using $1$\,nA (RuO$_2$) and $10$\,nA (Cernox) current at $18.15$\,Hz.

\begin{figure}[b]
\begin{center}
\includegraphics[width= 8.5cm]{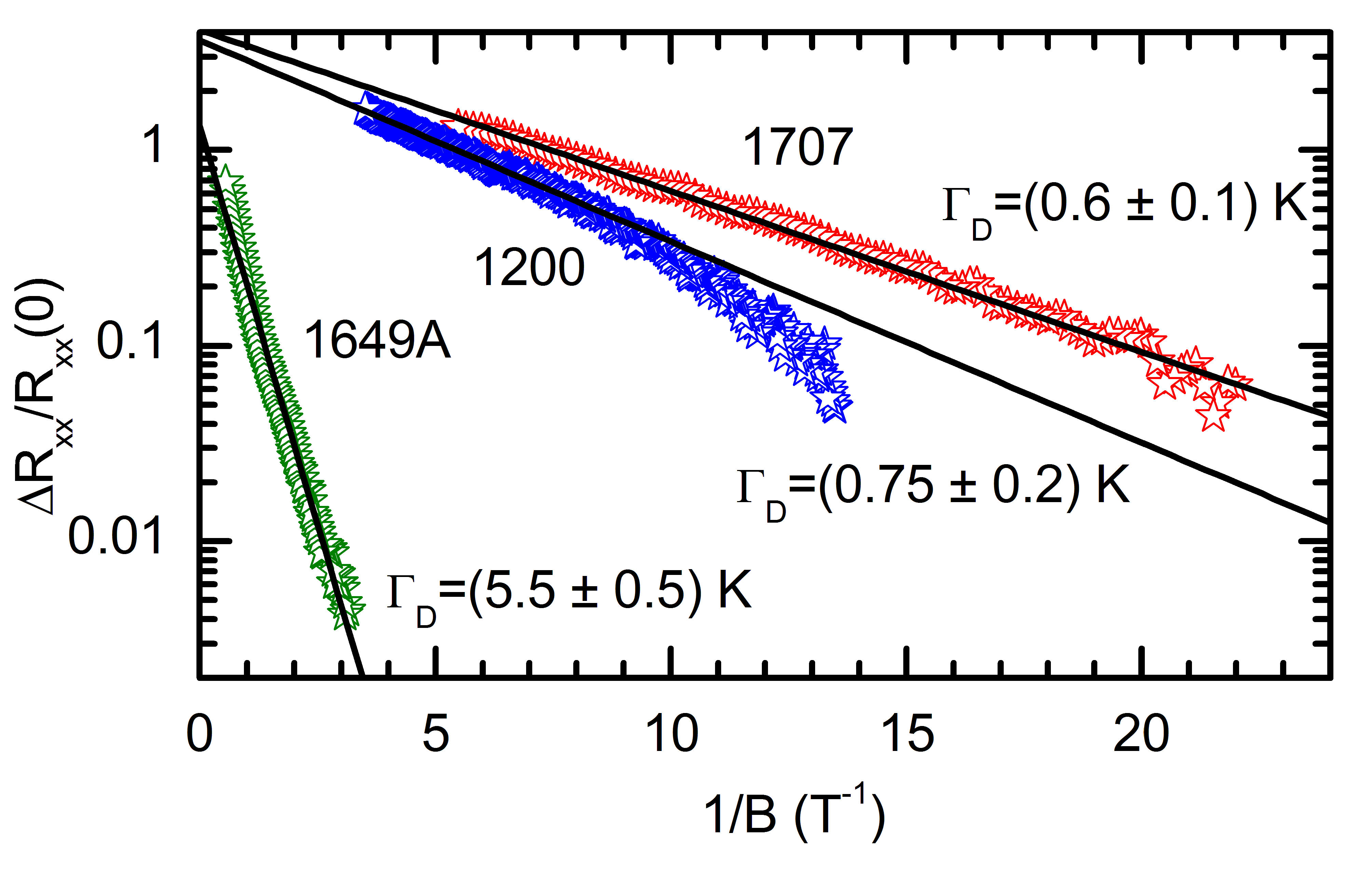}
\end{center}
\caption{(color online) Dingle plot of the amplitude of the oscillation $\Delta R_{xx}/R_{xx}(0)$ versus inverse
magnetic field for the three samples investigated. The solid lines are linear fits used to extract the Landau level
broadening.}\label{fig:AllSamplesDinglePlot}
 \end{figure}

The $R_{xx}$ versus magnetic field ($0-16$\,T) were measured with a sweep speed of $0.25$\,T/min, slow enough to avoid
(i) heating the fridge and (ii) shifting the high field quantum Hall data due to the $300$\,ms time constant of the
lockin amplifier. For the Dingle plots, particulary for the high mobility - high density samples, very slow sweeps were
performed ($0.005 - 0.05$\,T/min) to avoid damping the amplitude of the fast oscillations. We also corrected the data
for the $-27$\,mT remnant magnetic field of our superconducting coil which was determined from the characteristic zero
field weak localization cusp in $R_{xx}$. The critical magnetic field was determined, as in
reference\,[\onlinecite{Rigal1999}], by defining an arbitrary critical resistance $R_c = 10\,\Omega$, chosen to be
above the noise level in our system. For a current of $20$\,nA this corresponds to a voltage of $200$\,nV across the
voltage contacts of the sample.

\begin{table}\label{Tab:SampProp}
\caption{Summary of the sample parameters; the carrier density $n_s$, the transport mobility $\mu=e\tau_t/m^*$, the
Landau level broadening determined from the Dingle plot $\Gamma_D=\hbar/2\tau_q$, and the ratio of the transport and
quantum lifetimes $\tau_t/\tau_q$. Note the FWHM of the Landau levels is $2\Gamma_D$.}
\begin{center}
\begin{tabular}{ c | c   c  c c}

Sample  & $n_s$ (cm$^{-2}$) & $\mu$ (cm$^2$/Vs) & $\Gamma_D$ (K) & $\tau_t/\tau_q$\\
\hline
1649B & $7.28 \times 10^{11}$ & $1.0 \times 10^5$ & - & - \\

1649A & $7.28 \times 10^{11}$ & $7.5 \times 10^4$ & 5.5 & 5\\

1707 & $1.54 \times 10^{11}$ & $4.1 \times 10^6$ & 0.6 & 25\\

1200 & $7.49 \times 10^{11}$ & $2.5 \times 10^6$ & 0.75 & 20\\

\end{tabular}
\end{center}
\end{table}

All the GaAs/AlGaAs 2DEGs investigated were grown by molecular beam epitaxy using solid sources and modulation doping.
Sample 1649A is another piece of the same wafer as sample 1649B used in ref.\,[\onlinecite{Rigal1999}]. It has an
8.2\,nm wide GaAs quantum well. Sample 1707 is a standard high mobility GaAs/AlGaAs heterojunction.  Sample 1200 is a
13\,nm wide GaAs quantum well sandwiched between short period GaAs/AlAs superlattices which act as the barriers. The
two short period superlattices each have 60 periods of 4 ML of AlAs and 8 ML of GaAs. Carriers are introduced into the
central 13 nm GaAs quantum well by a single Si $\delta$-doping sheet with a concentration of $2.5 \times 10^{12}$
cm$^2$ placed in a GaAs layer of each short period superlattice. Carriers also occupy donor states associated with the
X conduction band minimum in the AlAs superlattice barriers. At low temperatures they are frozen out and do not
contribute to transport. However, they are very efficient at screening the potential fluctuations due to the $\delta$
doping giving unusually large mobilities for a high density 2DEG.\cite{Friedland1996,Faugeras2004} The Landau level
broadening was determined from a Dingle analysis of the low field data at mK temperatures. The Dingle plots of $\Delta
R_{xx}/R_{xx}(0)$ versus $1/B$ are shown in Figure\,\ref{fig:AllSamplesDinglePlot}. All samples show a linear behavior
over one to two orders of magnitude in resistance. The solid lines are the linear fits used to estimate the Landau
level broadening. A summary of all sample parameters can be found in Table\,\ref{Tab:SampProp}. The mobilities have
been calculated from the measured $T=0$ resistivity and the carrier density using $\mu = 1/n_se\rho$. To estimate the
ratio $\tau_t/\tau_q$ of the transport and the single particle (quantum) lifetime we have used $\mu=e\tau_t/m^*$ and
$\tau_q=\hbar/2\Gamma_D$. Values of $\tau_t/\tau_q$ of twenty are typical for high mobility 2DEGs due to the reduced
influence (factor of $1-\cos(\theta)$) of the small angle scattering on the transport lifetime. This is reduced to a
factor of roughly five for the low mobility quantum well sample 1649.

\section{Experimental results and discussion}\label{Sec:ExptResults}

\begin{figure}[b]
\begin{center}
\includegraphics[width= 8.5cm]{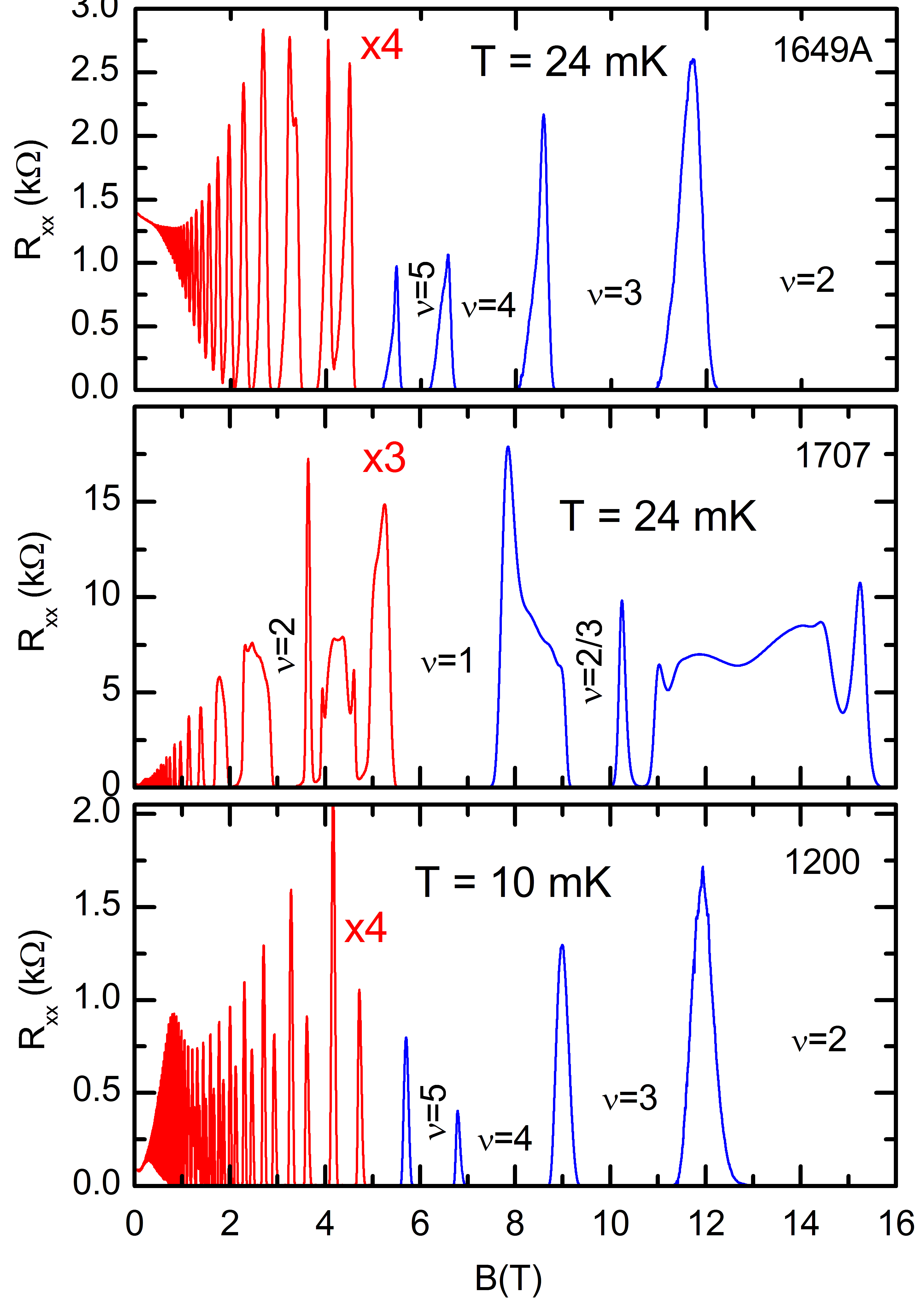}
\end{center}
\caption{(color online) The longitudinal resistance $R_{xx}$ as a function of magnetic field at low temperature for the
three samples investigated. The resistance in the low field regions have been multiplied by the indicated factors to
make the data at high filling factors more visible.}\label{fig:RxxAllSamples}
 \end{figure}

Figure\,\ref{fig:RxxAllSamples} shows the low temperature $R_{xx}$ versus magnetic field traces between 0 and 16\,T for
the three samples investigated. All samples show well defined integer quantum Hall states. Their properties are
summarized in Table\,\ref{Tab:SampProp}.

Sample 1649A is a low mobility - high density 2DEG with broad Landau levels with a high proportion of localized states
in the tails. It displays wide regions of dissipationless conduction in the vicinity of integer filling factors. Sample
1707 is a low density - high mobility 2DEG with narrow Landau levels and few localized states in the Landau level
tails. It displays relatively narrow regions of dissipationless conduction in the vicinity of integer filling factors.
1707 is a typical fractional quantum Hall sample which shows the standard series of fractions around filling factor
$\nu=1/2$ and $\nu=3/2$, notably a well developed $\nu=2/3$ state with a wide region of dissipationless conductance.
Sample 1200 shows a very similar high field $R_{xx}(B)$ trace as sample 1649 due to the similar densities and
proportion of localized states.

In order to determine the $B_c(T)$ phase diagram we have made an extremely detailed temperature dependence of
$R_{xx}(B)$ for each sample. Below we present the results sample by sample together with the fits to the thermal
activation model. Global conclusions, comparing the behavior of the three samples in order to probe the validity of the
assumptions made in the thermal activation model, will be drawn in the discussion section. We will see that a self
consistent picture emerges, with clear evidence that the mobility edge actually moves if spin splitting at even filling
factors is suppressed, or if the Landau level broadening or overlap changes.

\subsection{Sample 1649A - high density - large disorder - low mobility}

\begin{figure}[tb]
\begin{center}
\includegraphics[width= 8.5cm]{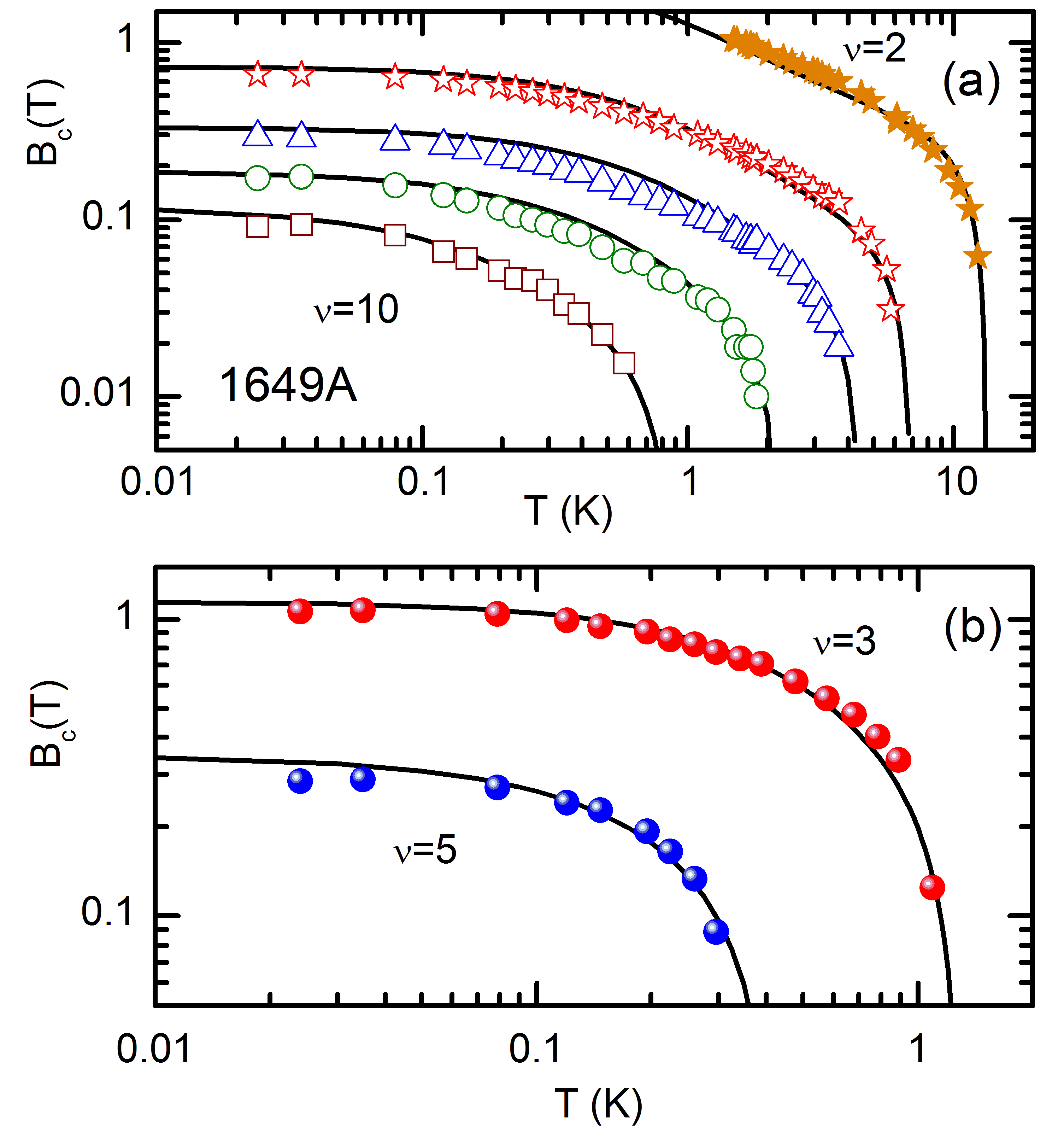}
\end{center}
\caption{(color online) Critical magnetic field versus temperature plotted as log log for sample 1649A for (a) various
even filling factors and (b) various odd filling factors. The solid lines are calculated using the thermal activation
model as described in the text.}\label{fig:All1649Armelle}
 \end{figure}

Samples 1649A is another piece of the same wafer as 1649B ($B_c(T)$ data presented in figure\,\ref{fig:Even1649Barry}).
We have completely repeated the measurements to have a full data set for both odd and even filling factors and to have
an estimate of the Landau level width from a Dingle analysis (see section\,\ref{Sec:ExptTech}). In this sample we have
access to even filling factors $\nu=2,4,6,8,10$, although the high field extremity of $\nu=2$ is only in field range
for high temperatures ($T>1$K). We have well developed dissipationless regions for odd filling factors $\nu=3,5$, with
inevitably filling factor $\nu=1$ out of the available field range. The $B_c(T)$ phase diagram is plotted in
figure\,\ref{fig:All1649Armelle} for both odd and even filling factors. For even filling factors the phase diagram is
quasi identical to that measured previously for 1649B, although the values of $B_c$ found are slightly lower.

Fitting to the lowest even integer filling factor for which we have data over the full temperature range ($\nu=4$), we
obtain $\Gamma=3.4K$, similar to the value found for 1649B, but slightly smaller than the broadening, determined from
the Dingle analysis, of $\Gamma_D=5.5$K suggesting a narrowing of the Landau levels at high field. The value of $B_c(T
\rightarrow 0)$ fixes $\Gamma_{dl} = 1.1\,\Gamma$, slightly larger than the value found for 1649B ($0.9 \Gamma$)
indicating that 1649A has fewer localized states which leads to the slightly lower values of $B_c$. An excellent fit to
the $\nu=4$ data is obtained using the same thermal broadening parameter $\alpha=1$ (solid line in
figure\,\ref{fig:All1649Armelle}(a)). The cyclotron gap was calculated using an effective mass $m^*=0.072 m_e$ slightly
larger than the band edge mass in GaAs due to non parabolicity.\cite{Raymond1979} The thermal activation model is then
used to calculate the $B_c(T)$ phase diagram for all the even integer filling factor with \emph{no adjustable
parameters}. The agreement with experiment is remarkable confirming our simple model in which the Landau level width is
independent of magnetic field, with a mobility edge which does not move (\emph{i.e.} $\Gamma_{dl}/\Gamma$ is constant,
independent of the filling factor).

For odd filling factors we fit to the $\nu=3$ data, fixing all parameters as for the even filling factors (see
Table\,\ref{Tab:FitParam}), with the exception of $\Gamma_{dl}=0.3 \Gamma$ and an effective g-factor for the many body
enhanced spin gap $g^*_{eff}=2.4$. The fit is again excellent (solid lines figure\,\ref{fig:All1649Armelle}(b)) for
both $\nu=3$ and $5$. This demonstrates that as for the even filling factors the Landau level broadening and the
position of the mobility edge is independent of the filling factor. $\Gamma_{dl}$ is approximately a factor of four
smaller than for the even integer case demonstrating that opening of the spin gap causes the mobility edge to shift
significantly towards the center of the Landau level creating many more localized states in the tail of the Landau
level. On the other hand $\Gamma$ is unchanged for odd and even filling factors showing that the opening of the spin
gap does not affect the Landau level broadening.

\subsection{Sample 1707 - low density - low disorder - large mobility}

\begin{figure}[]
\begin{center}
\includegraphics[width= 8.5cm]{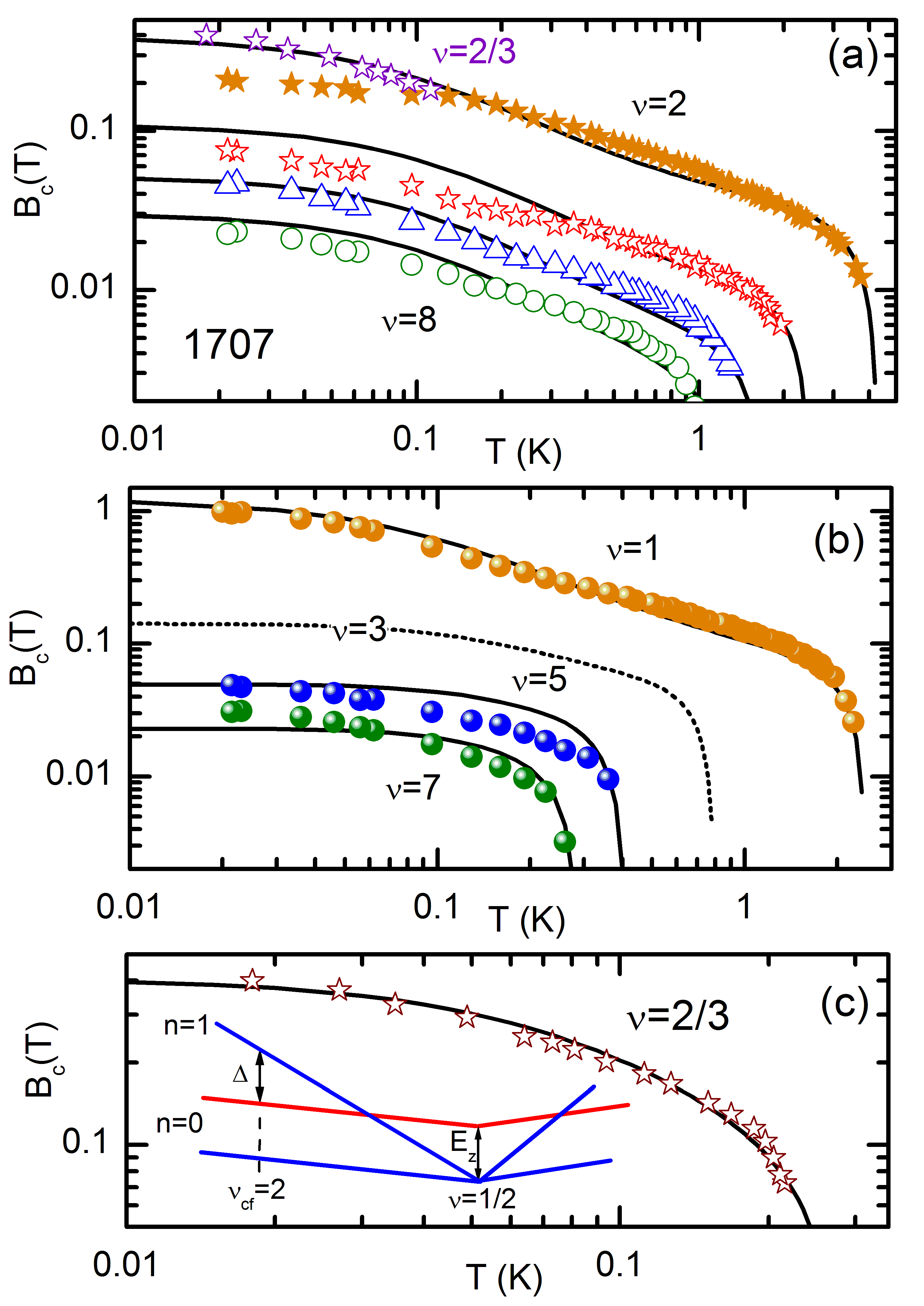}
\end{center}
\caption{(color online) Critical magnetic field versus temperature plotted as log log for sample 1707 for (a) various
even, (b) odd  and (c) fractional filling factors. The solid lines are calculated using the thermal activation model as
described in the text. The inset shows a schematic of the composite Fermion Landau levels around $\nu=1/2$. In (a) we
plot also the data of $\nu=2/3$ which is the composite Fermion filling factor $\nu_{cf}=2$. The agreement with electron
filling factor $\nu=2$ data is remarkable. }\label{fig:All1707}
 \end{figure}

In sample 1707 we have access to even filling factor $\nu=2,4,6,8$ and odd filling factors $\nu=1,5,7$. For some reason
in this sample the $\nu=3$ minimum in $R_{xx}$ has an asymmetric shape, lifting off prematurely on the high field side
and never actually falls below the $R_c = 10\Omega$ cut off used to determine $B_c$. In addition, 1707 has several
prominent fractions, of which only $\nu=2/3$ is fully developed, with $R_{xx}$ falling below the $R_c = 10\Omega$ cut
off. Unfortunately, $\nu=1/3$ remains out of our field range, even at higher temperatures. The $B_c(T)$ phase diagram
is plotted in figure\,\ref{fig:All1707} for both odd, even and fractional filling factors. Compared to 1649, the even
integer filling factors have relatively small values of $B_c$ at low temperature reflecting the lower magnetic field at
which they occur (due to the lower carrier density) and the greatly reduced disorder in this sample.

While we could adopt the same procedure as for sample 1649, fitting to $\nu=2$ to obtain the correct parameters, and
then generating curves for all the other filling factors without any adjustable parameters, it turns out that this is
not the best approach. There is a suspicion that the even filling factors are fragile in this sample, and the low
temperature values of $B_c$ may be lower than they should be for whatever reason, \emph{e.g.} despite the small current
used the Hall electric field may shift the mobility edge. In addition, the parameters obtained by fitting to the
$\nu=2$ data do not fit extremely well the other filling factors; notably the predicted low temperature values of $B_c$
are \emph{lower} than the measured values.

Fitting to the $\nu=1$ data, we obtain a Landau level broadening $\Gamma=0.23$\,K, lower than the value obtained from a
Dingle analysis $\Gamma_D=0.6$\,K suggesting again that the Landau level width at high field is considerably reduced
compared to the low field value. We stress, that the width of the low temperature phase of $B_c(T)$ phase diagram is
very sensitive to the width of the Landau level so that $\Gamma$ is determined quite precisely. The experimental value
of $B_c(T \rightarrow 0)$ is correctly reproduced with $\Gamma_{dl} = 1.5 \Gamma$ which is much larger than the value
($0.3 \Gamma$) obtained for odd filling factors in 1649A; the fraction of localized states is much lower in sample
1707. Thermal broadening effects are also very small with $\alpha=0.3$. As for 1649 the agreement between the
predictions of the thermal activation model and the $\nu=1$ data is excellent over nearly two orders of magnitude in
temperature and magnetic field. Using the same parameter set, reasonable agreement is obtained for filling factors
$\nu=5$ and $7$. The size of the many body spin gap has been calculated using a filling factor dependent effective
g-factor $g_{eff}^* = 2.5 - 7.5$. to correctly reproduce the observed temperature dependence of $B_c(T)$.

The predicted $B_c(T)$ for even filling factors is then calculated using exactly the same parameter set as for the odd
filling factors (see Table\,\ref{Tab:FitParam}). In this low density sample, the cyclotron gap was calculated using an
effective mass $m^*=0.067$ which is the band edge mass in GaAs. The agreement between model (solid lines) and the data
is good with some deviation at low temperatures with the measured values being too low, especially for $\nu=2$. As for
sample 1649A, odd and even filling factors can be fitted with the same Landau level broadening $\Gamma$. However, in
stark contrast, both sets of filling factors can be fitted without moving the mobility edge \emph{e.g.} the same value
of $\Gamma_{dl}/\Gamma$. This would be consistent with the small single particle spin gap being open, in the absence of
exchange interactions, at spin unpolarized even filling factors. This is reasonable since in 1707, the single particle
Zeeman energy $E_z \simeq 1$\,K (at $\nu=2$) is much greater than the Landau level width $\Gamma=0.23$\,K.

The idea that the experimental values of $B_c(T)$ are too low at low temperatures is comforted by the data for
$\nu=2/3$ which corresponds to composite Fermion filling factor $\nu_{cf}=2$.\cite{Jain1989} In this picture, the many
body fractional quantum Hall effect of electrons is mapped to the integer quantum Hall effect of non interacting
composite Fermion quasi particles. Composite Fermions are formed by attaching two fictitious magnetic flux quantum,
antiparallel to the applied magnetic field, to each electron. In a mean field picture, composite Fermions move in an
effective magnetic field $B^* = B - B_{1/2}$ where $B_{1/2}$ is the magnetic field at electron filling factor
$\nu=1/2$. The low temperature $\nu=2/3$ data, plotted in figure\,\ref{fig:All1707}(a) lies above the $\nu=2$ data,
indicating that the composite Fermion $\nu_{cf}=2$ state is more robust than its electron counter part, and is in
excellent agreement with the prediction of the activation model for electron filling factor $\nu=2$.

Finally, we apply the activation model to the $\nu=2/3$ fractional state (see figure\,\ref{fig:All1707}(c)). In the
framework of the composite Fermion model we treat this state as an effective $\nu_{cf}=2$ state. We can fit to the data
with $\Gamma = 0.23$ confirming a previous report, based on a Dingle analysis, that the composite Fermions and
electrons have identical Landau level broadening.\cite{Leadley1994} The value of $B_c(T \rightarrow 0)$ fixes
$\Gamma_{dl} = 0.9 \Gamma$. Imposing $g^*=-0.44$ for the non interacting composite Fermions an excellent fit is
obtained with $m^*_{cf}=0.8$. Leadley \emph{et al.} reported that the composite Fermion mass varies as a function of
the effective magnetic field $B^*$.\cite{Leadley1994} In our sample $\nu=2/3$ occurs at $B^*\simeq3.2$\,T (the same
magnetic field as $\nu=2$). The literature value for composite Fermion mass at $B^*$ is  $m^*_{cf}\simeq0.7-1.0$, in
reasonable agreement with out value.\cite{Leadley1994,Du1994} For electrons, in the activation model the effective mass
essentially controls the gap at even filling factors (corrections due to the single particle Zeeman energy are
negligible). For composite Fermions the situation is complicated by the large mass, which reduces the cyclotron gap,
and by the effectively larger single particle Zeeman energy, which depends on the total magnetic field ($3B_f/2$)
rather than $B^*=B_f/2$. For our parameter set we have at $\nu_{cf}=2$, a cyclotron energy $\hbar\omega_c \simeq
5.4$\,K and a Zeeman energy $E_z = 2.9$\,K. As $\hbar\omega_c > E_z$ this implies that the ground state is spin
unpolarized (the spin up and down $n=0$ composite Fermion spin Landau levels are occupied) and excitations involve a
spin flip to the $n=1$ composite Fermion Landau level. The situation, with the composite Fermion Landau levels is shown
schematically in the inset of figure\,\ref{fig:All1707}(c). The energy gap for the excitation is $\Delta =
\hbar\omega_c - E_z$, which depends on both the effective mass and the composite Fermion g-factor.

\subsection{Sample 1200 - high density - large disorder - high mobility}

\begin{figure}[]
\begin{center}
\includegraphics[width= 8.5cm]{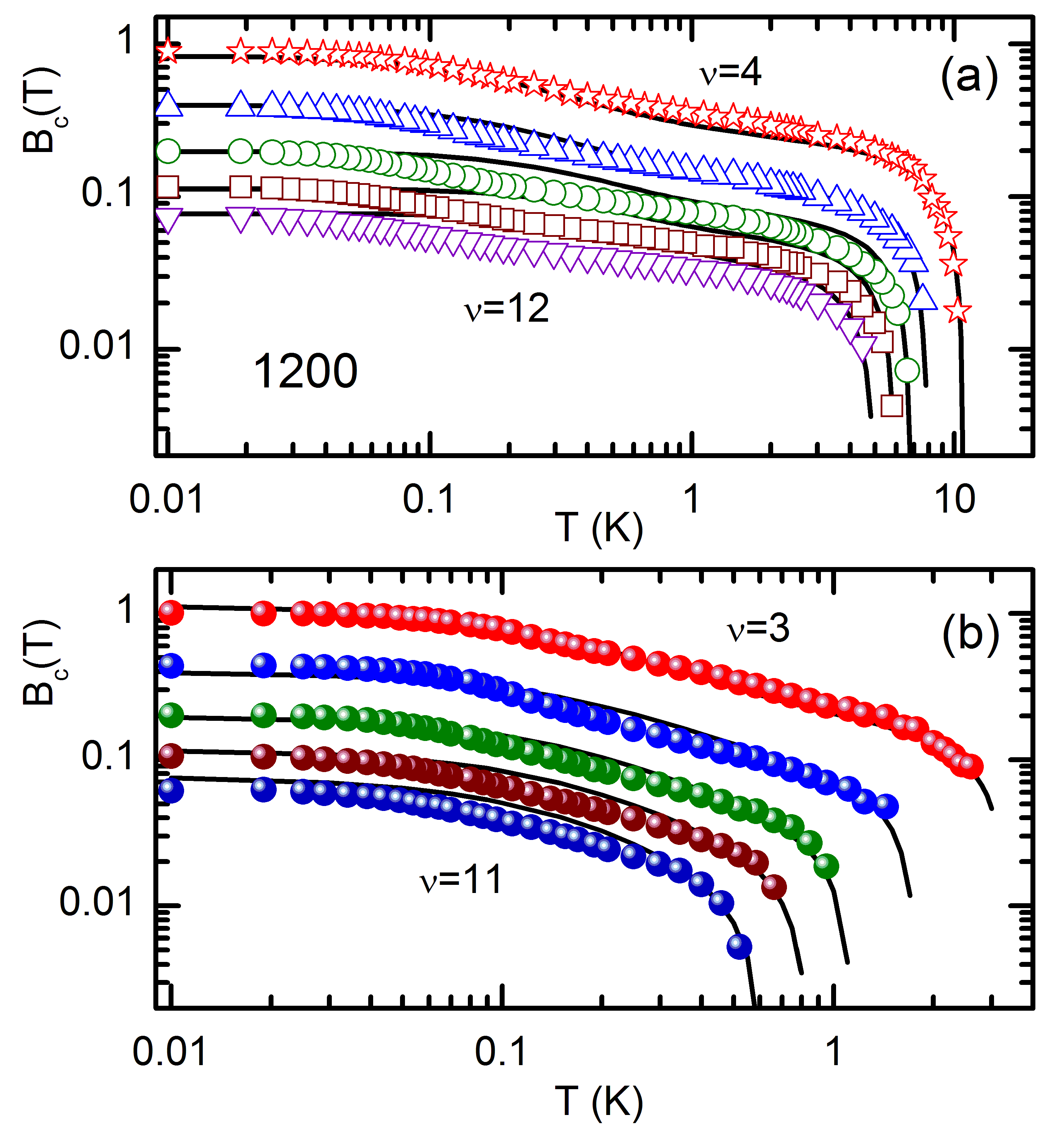}
\end{center}
\caption{(color online) Critical magnetic field versus temperature plotted as log log for sample 1200 for (a) various
even filling factors and (b) various odd filling factors. The solid lines are calculated using the thermal activation
model as described in the text.}\label{fig:All1200}
 \end{figure}

Sample 1200 is a rather unusual 2DEG due essentially to the super lattice barriers. It has a high carrier density
together with a high mobility, but also a rather large disorder in the sense that it has narrow Landau levels but a
large proportion of localized states. This combination gives rise to a large number of accessible odd ($\nu=3 - 11$)
and even ($\nu=4-12$) filling factors. Here we have limited the analysis of even filling factors to states which have
$B_c(T \rightarrow 0)>50$\,mT. The $B_c(T)$ phase diagram is plotted in figure\,\ref{fig:All1200} for odd, and even
filling factors. Both show robust dissipationless states with values of $B_c$ at low temperatures comparable to the
highly disordered sample 1649.

As for sample 1707, we start by fitting to the lowest odd filling factor $\nu=3$. The width of the low temperature
phase is best described with a Landau level broadening of $\Gamma=0.9$\,K, which in contrast to the other samples is
actually larger than the broadening $\Gamma_D=0.75$\,K extracted from a Dingle analysis of the low field oscillations.
The value of $B_c(T \rightarrow 0)$ fixes the position of the mobility edge with $\Gamma_{dl} = 0.58 \Gamma$. We use a
cyclotron mass $m^*=0.072 m_e$ to correct for non parabolicity, a thermal broadening parameter $\alpha=1.25$, and the
many body enhanced spin gap is calculated using an effective g-factor $g^*_{eff}=7.0$. The agreement between the model
(solid line) and experiment is excellent. The curves for the other filling factors can then be generated with
$g^*_{eff}=7.0-5.0$ as the only adjustable parameter. As for the other samples the data for all filling factors is well
reproduced over $2-3$ orders of magnitude in magnetic field and temperature. This demonstrates that for the odd filling
factors the Landau level broadening and the position of the mobility edge are independent of filling factor. Note that
this conclusion, which can be drawn from the $B_c(T \rightarrow 0)$ data alone, is independent of the value of
$g^*_{eff}$ used, which simply improves the fit at intermediate temperatures.

\begin{table*}[]\label{Tab:FitParam}
\caption{Summary of the parameters used in the thermal activation model; the effective mass $m^*$ in GaAs corrected for
non parabolicity in the high density samples, the factor $\alpha$ used to include thermal broadening, the $T=0$ width
of the Landau levels $\Gamma$, the ratio of the width of the localized and delocalized state $\Gamma_{dl}/\Gamma$ for
even, odd and fractional filling factors, the Land\'e g-factor $g^*$ for GaAs, and the effective g-factor $g^*_{eff}$
to describe the exchange enhanced spin gap at odd filling factors, and for fractional filling factors the composite
fermi effective mass and g-factor.}
\begin{center}
\begin{tabular}{ c |c c  c  c | c  c | c c | c c c}
  & All & & & & Even & & Odd & & Fractional & &\\
Sample & $B_f$ (T) & $m^*(m_e)$  & $\alpha$ & $\Gamma$ (K)& $\Gamma_{dl}/\Gamma$ & $g^*$ & $\Gamma_{dl}/\Gamma$ & $|g^*_{eff}|$ & $\Gamma_{dl}/\Gamma$ & $m^*_{cf}(m_e)$ & $g^*_{cf}$\\
\hline
1649B & 30.15 & 0.072 & $1.0 \pm 0.1$ & $3.0 \pm 0.3$  & $0.9 \pm 0.1$ & -0.44 & - & - & - & - & -\\

1649A  & 30.15 & 0.072 & $1.0 \pm 0.1$ & $3.4 \pm 0.3$ & $1.1 \pm 0.1$ & -0.44 & $0.3 \pm 0.05$ & 2.4 & - & - & -\\

1707 & 6.37 & 0.067  & $0.3 \pm 0.05$ & $0.23 \pm 0.05$ & $1.5 \pm 0.1$ & -0.44 & $1.5 \pm 0.1$ & $2.5-7.5$ & $0.9 \pm 0.1$ & $0.8 \pm 0.1$ & -0.44\\

1200 & 31.00 & 0.072 & $1.25 \pm 0.1$ & $0.9 \pm 0.1$ & $1.3 - 0.58$ & -0.44 & $0.58 \pm 0.05$ & $5.0-7.0$ & - & - & -\\

\hline
\end{tabular}
\end{center}
\end{table*}

To fit the even filling factors we start with an identical parameter set as for the odd filling factors, with of course
the exception that the spin gap is calculated with the single particle g-factor $g^*=-0.44$. An excellent fit is
obtained for $\nu=4$ and $\nu=6$. For lower filling factors the $B_c(T \rightarrow 0)$ data can only be reproduced by
assuming that the mobility edge is shifting outwards into the tail of the Landau level. With this assumption,
reasonable fits are obtained using $\Gamma_{dl} = 0.58, 0.58, 0.9, 1.24,$ and $1.3 \Gamma$ for filling factors $\nu=4,
6, 8 ,10, 12$ respectively. The agreement is not as good as for the other samples; while the values of $B_c(T
\rightarrow 0)$ are perfectly reproduced, the predicted values of $B_c$ at intermediate temperatures is too large
indicating that the Landau level broadening decreases at filling factors above $\nu=6$. This discrepancy is also
visible for the higher odd filling factors. The decrease of $\Gamma$ in this sample, possibly linked to the changing
ratio of the magnetic length and the characteristic length scale of the disorder potential, would necessarily cause the
mobility edge to shift outwards, as required to fit to the $B_c(T \rightarrow 0)$ data.

If the single particle spin gap was open at $\nu=4$ and $\nu=6$, before progressively closing for higher filling
factors this would also cause the mobility edge to shift out. This is plausible as the single particle Zeeman energy
$E_z \simeq 1.2$\,K at $\nu=8$ is comparable to the Landau level broadening $\Gamma=0.9$\,K. However, if the Landau
level broadening is decreased then the spin gap should remain open for the higher filling factors.

\subsection{Discussion - validity and limits of the model}

The picture which emerges from fitting the $B_c(T)$ phase diagram for odd and even filling factors for the three
samples above can be summarized as follows. The Landau level broadening and the position of the mobility edge within
the Landau level are independent of the magnetic field (Landau level index) provided the overlap of the spin up and
spin down sub levels is not changing. Under such conditions the simple intra Landau level thermal activation model
provides and accurate description of the $B_c(T)$ phase diagram for all filling factors with a single parameter set. As
the single particle Zeeman energy is small, paradoxically odd filling factors, with their large exchange enhanced spin
gaps fulfill this condition for all of the investigated samples. The behavior at even filling factors, for which there
is no enhanced spin gap, depends on the size of the single particle Zeeman energy compared to the Landau level
broadening. In sample 1707 with it extremely narrow Landau levels, to a first approximation the spin gap remains fully
open at even integer filling factors, and both odd and even filling factors can be fitted with the same position of the
mobility edge. In sample 1649, the Landau levels are broad and the spin gap remains closed at even filling factors. Two
different positions of the mobility edge are required for even and odd filling factors. Finally, in sample 1200, while
the odd filling factors are well behaved and can be fitted with a single position for the mobility edge, for even
filling factors, the spin gap is opening at lower filling factors and the mobility edge shifts closer to the Landau
level center, finally reaching the same value as for odd filling factors.

\begin{figure}[]
\begin{center}
\includegraphics[width= 8.5cm]{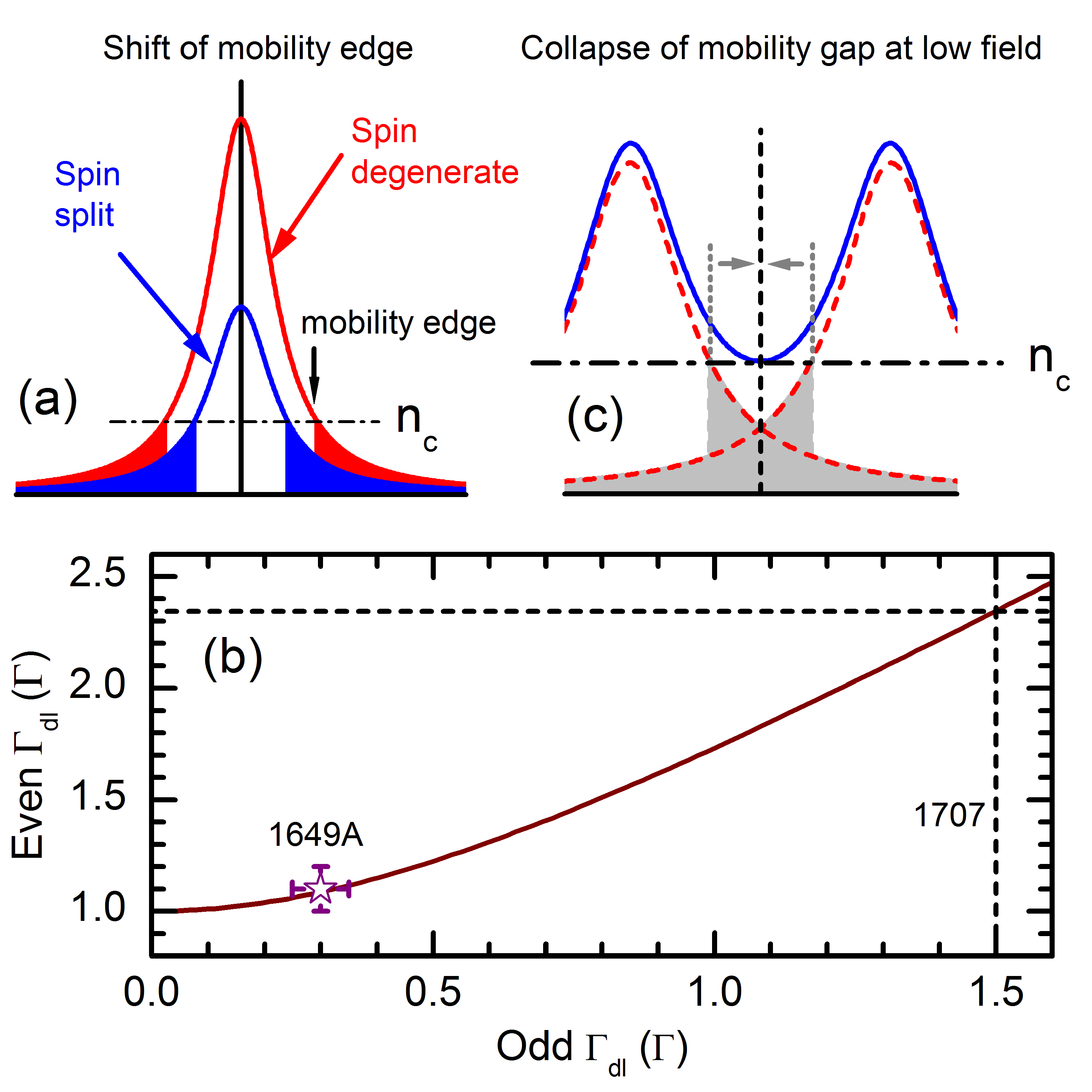}
\end{center}
\caption{(color online) (a) Schematic of the density of states of a Landau level with and without spin splitting. To
maintain the same density of states, the mobility edge has to move towards the center of the Landau level when the spin
degeneracy is lifted. (b) The solid line shows the predicted position of the model edge when the spin gap is closed
versus the position of the mobility edge when the spin gap is open. The symbols shows the experimental values for
sample 1649A. (c) Schematic showing two Landau levels (broken lines) strongly overlapping at low magnetic field. The
solid line shows the total density of states while the grey areas show localized states (mobility edge) for non
overlapping Landau levels. To maintain the same density of states (without overlap) the mobility edge has to shift
towards the tail of the Landau level, eventually causing the mobility gap to collapse.}\label{fig:MobEdgeMove}
 \end{figure}

At the critical magnetic field, breakdown occurs due to the onset of variable range hopping in the tail of the Landau
level. At $T=0$, the mobility edge corresponds to the critical density of states at the Fermi level. In analogy to the
quasi elastic inter Landau level scattering (QUILLS) model,\cite{Eaves1986} in this case in the absence of electric
field and within the same Landau level, we can estimate the number of states in the Landau level which have sufficient
wave function overlap for tunneling. To have sufficient overlap, the states have to lie with a circle of radius $2
A_n$, where $A_n = (2n + 1)^{1/2} \ell_B$ is the classical turning points of the simple harmonic oscillator,
$\ell_B=(h/eB)^{1/2}$ is the magnetic length, and $n$ is the orbital quantum number. The Landau level degeneracy can be
written as $2\pi/\ell_B^2$ so that the number of states which can participate at an energy $E$ (measured from the
center of the Landau level) is given by
\begin{equation}
n_p(E) = \lambda \frac{8 \pi^2 (2n+1)}{ \pi \Gamma (1 + E^2/\Gamma^2)},\nonumber
\end{equation}
where $\lambda=1$ if the Landau level is spin split or $\lambda=2$ if the spin gap is closed. The magnetic lengths in
the Landau level degeneracy and $A_n$ cancel leaving only the prefactor $(2n+1)$ which describes the increased
delocalization of the higher energy simple harmonic oscillator wave functions. Thus, in contrast with the experimental
observation that the mobility edge remains fixed, the number of states close enough to tunnel is predicted to depend on
the Landau level index (magnetic field). In this picture, the mobility edge, $\Gamma_{dl}$, corresponds to the hopping
threshold, when the critical number of states per unit area \emph{at the same energy}, ($n_c = n_p(\Gamma_{dl})$),
which can tunnel is achieved.

When the single particle spin gap closes, $\lambda \rightarrow 2$ and the mobility edge will have to move to a new
position further out ($\Gamma_{dl}^{'}$) into the tail of the Lorentzian to maintain the critical density (see
schematic in figure\,\ref{fig:MobEdgeMove}(a)). The relation between the two values can be written as
\begin{equation}
\Gamma_{dl}^{'} = \sqrt{\Gamma^2 + 2 \Gamma_{dl}^2}.\nonumber
\end{equation}
In figure\,\ref{fig:MobEdgeMove}(b) we plot (solid line) the predicted position of the mobility edge $\Gamma_{dl}^{'}$
when the spin gap is closed versus $\Gamma_{dl}$. The measured value for sample 1649A (symbol), is in good agreement
with the simple model. We cannot compare for the other samples since either the spin gap remains open (1707) or the
mobility edge is moving continuously (1200).

Note also, that comparing the values of $\Gamma_{dl}/\Gamma$ (see Table\,\ref{Tab:FitParam}) for even filling factors,
gives the misleading impression that all samples have a similar disorder. This is due to the spin gap remaining open at
even filling factors in sample 1707. Comparing $\Gamma_{dl}/\Gamma$ for odd filling factors show that the width of the
delocalized states is $3-5$ times larger for sample 1707. The dashed lines in figure\,\ref{fig:MobEdgeMove}(b)
indicates the expected value, for sample 1707, of $\Gamma_{dl}/\Gamma \simeq 2.35$ if the spin gap was closed at even
filling factors.

\begin{figure}[]
\begin{center}
\includegraphics[width= 8.5cm]{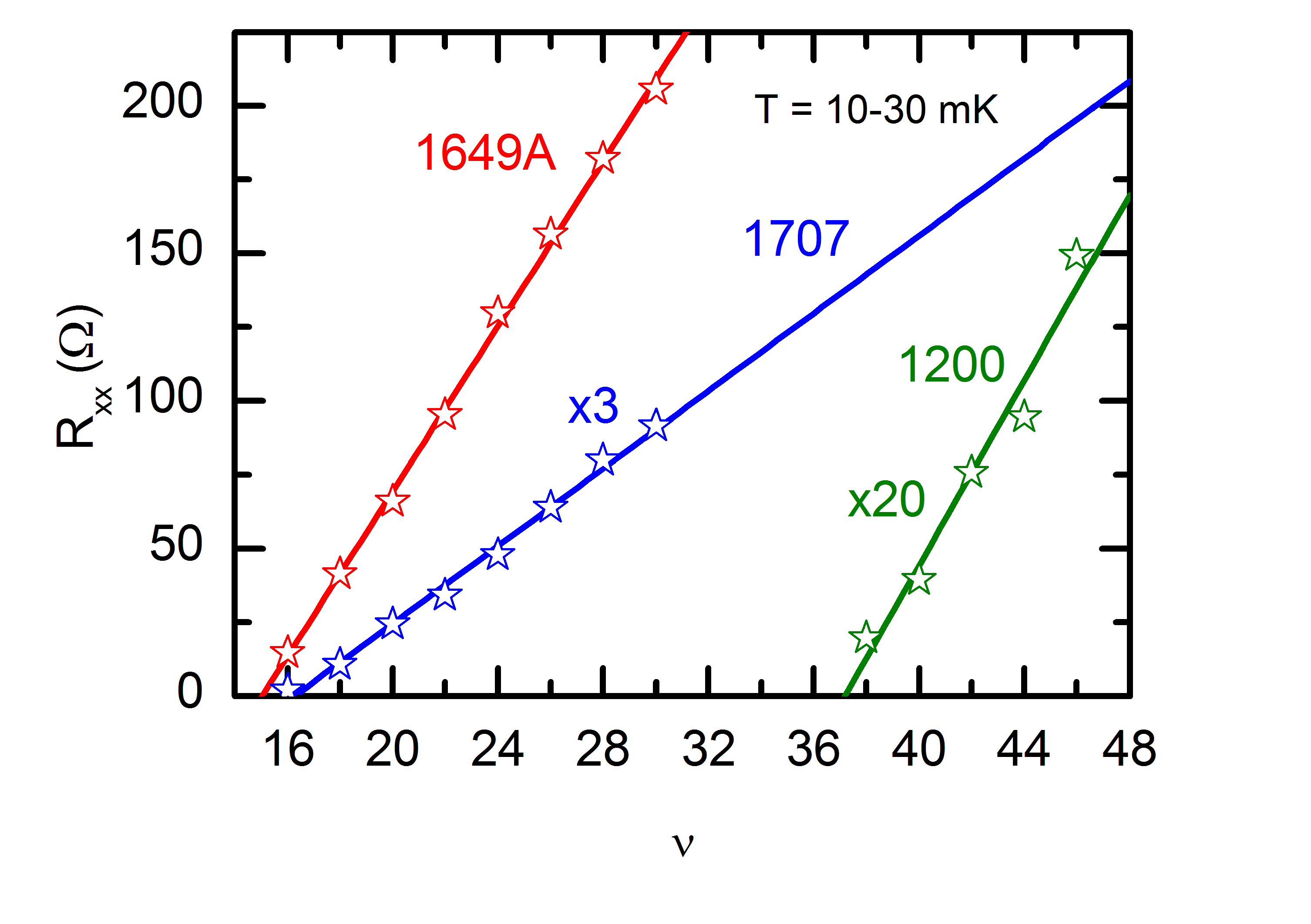}
\end{center}
\caption{(color online) Value of the low temperature resistance versus even filling factor for the three samples
investigated. The solid lines are linear fits. The intercept with the horizontal axis can be used to estimate the
filling factor $\nu_0$ at which the dissipationless state disappears at $T=0$.}\label{fig:RcNu0AllSamples}
 \end{figure}

\begin{figure}[tb]
\begin{center}
\includegraphics[width= 8.5cm]{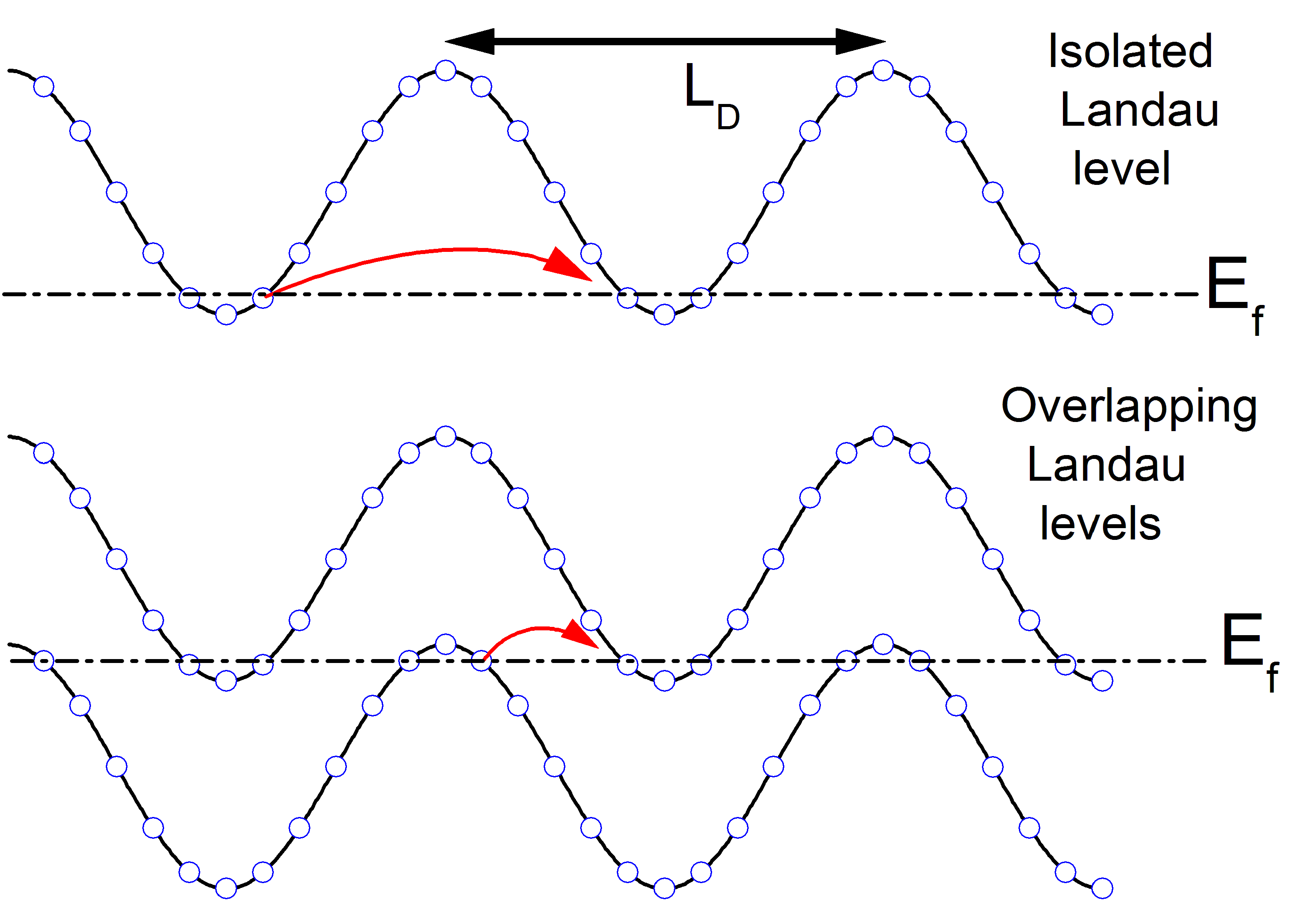}
\end{center}
\caption{(color online) Schematic of the spatial, disorder induced, potential fluctuations in a Landau level. The
potential fluctuations have a characteristic length scale $\ell_D$ which controls the hopping distance required to find
another state at the same energy (\emph{i.e.} at the Fermi energy $E_f$). If the Landau levels overlap this distance is
roughly halved.}\label{fig:Disorder}
 \end{figure}

Finally, it is interesting to consider the predictions of the activation model in the low field limit for even filling
factors. In figure\,\ref{fig:RcNu0AllSamples} the resistance at even filling factors at low temperatures ($T
\rightarrow 0$) is plotted as a function of filling factor. The dependence is approximately linear and the intercept
with the horizontal axis can be used to estimate the filling factor at which the conductance ceases to be
dissipationless, $\nu_0=14, 16$ and $38$ for samples 1649A, 1707 and 1200 respectively. Assuming that
$\Gamma_{dl}/\Gamma$ is independent of filling factor, we would naively expect that at $T=0$ the dissipationless
conductance would disappear at filling factor $\nu_0$ when $2\Gamma_{dl}=E_c^{\nu=1}/\nu_0$ where $E_c^{\nu=1} = \hbar
e B_f/m^*$ is the cyclotron energy at $\nu=1$. This prediction is clearly wrong, for example for sample 1649A, using
the parameters from Table\,\ref{Tab:FitParam} gives $\nu_0 \simeq 74$ while the conductance is no longer
dissipationless for $\nu \geq 14$. What is not included in the model is the movement of the mobility edge as the
adjacent Landau levels start to overlap; the mobility edge has to shift out so that the overlapping density of states
at the mobility edge maintains the same value of $n_c$. The mobility gap finally collapses, when the combined density
of states at the center of the cyclotron gap equals $n_c$ (see figure\,\ref{fig:MobEdgeMove}(c)). The condition for
this can be written,
\begin{equation}\label{Eq:nu_0}
\nu_0 = \frac{E_c^{\nu=1}}{2 \sqrt{\Gamma^2 + 2 \Gamma_{dl}^2}}.
\end{equation}
This gives $\nu_0 \simeq 58$ for sample 1649A which is still too large. However, the above expression considers only
states from the Landau levels immediately above and below the Fermi level which is not a good approximation. Taking
into account contributions from all Landau levels the condition can be written as
\begin{equation}
\frac{2}{\pi\Gamma(1 + \Gamma_{dl}^2/\Gamma^2)} = \sum_{n=0}^{\infty} \frac{2}{\pi\Gamma \left(1 +
\left(\frac{E_c^{\nu=1}}{2 \nu_0 \Gamma}\right)^2\left( \nu_0 - 1 -2n\right)^2\right)}.\nonumber
\end{equation}
The L.H.S. is the density of states (in units of $eB/h$) at the mobility edge of an isolated Landau level and the
R.H.S. is the sum of the density of states in the middle of the cyclotron gap (at the Fermi energy) for filling factor
$\nu_0$. If only the (identical) terms with $n=\nu_0/2$ and $n=-1 + \nu_0/2$ are retained, the expression can be
rearranged to obtain the simplified expression of Equation\,\ref{Eq:nu_0}. The infinite sum converges rapidly and for
our purposes it is more than enough to take into account the first $200$ Landau levels. Using the parameters from
Table\,\ref{Tab:FitParam}, the predicted values of $\nu_0 \simeq 38$ (1649A), $\nu_0 \simeq 68$ (1707) and , $\nu_0
\simeq 132$ (1200) which are all significantly larger than the experimental values. This suggests that the overlap of
multiple Landau levels causes the mobility edge to move out more rapidly than predicted by our simple model. All Landau
levels see the same potential fluctuations, so that within a Landau level the minimum hop distance is given by the
characteristic length of the disorder potential $\ell_D$. The overlap between the high energy tail of one Landau level,
with the low energy tail of another, will halve the minimum hopping distance to $\ell_D /2$ (see
figure\,\ref{fig:Disorder}). This is not included in the simple model; the reduced hopping distance will reduce the
critical density of states required for hopping and cause the mobility edge to move rapidly outwards, precipitating the
collapse of the mobility gap at low fields.

\section{Implications for scaling theory}\label{Sec:ScalingTheory}

In the pioneering work of Pruisken, a scaling theory of the plateau plateau (PP) or plateau insulator (PI) transitions
in the QHE regime was developed, using renormalization group theory (RGT), to describe the problem of a quantum phase
transition which is too complicated to be solved from first principles.\cite{Pruisken1988} The theory predicts a
universal behavior, characteristic of a quantum phase transition, in which the width of the peak in $R_{xx}$ scales as
$\Delta B \propto T^{\kappa}$ where $\kappa$ is universal. The theory predicts no particular value for $\kappa$ and
experiment, from a log log plot of $1/\Delta B$ versus $T$, initially concluded that $\kappa=0.42$.\cite{Wei1988}
Subsequent work, including measurements on the same sample used in Ref.[\onlinecite{Wei1988}], found a larger value of
$\kappa=0.58$.\cite{Ponomarenko2004,Lang2005,Ponomarenko2005,Visser2006,Lang2007} In the original experiment, Wei
\emph{et al.} defined the peak width $\Delta B \propto T^{\kappa}$, whereas, later work defined the width in terms of
filling factor $\Delta \nu \propto T^{\kappa}$. As the filling factor is inversely proportional to the magnetic field
the temperature dependence of $\Delta B$ and $\Delta \nu$ can be somewhat different; it is not obvious that if one
follows a power law, that the other will also follow the same power law.\cite{Ponomarenko2005}

\begin{figure}[]
\begin{center}
\includegraphics[width= 8.5cm]{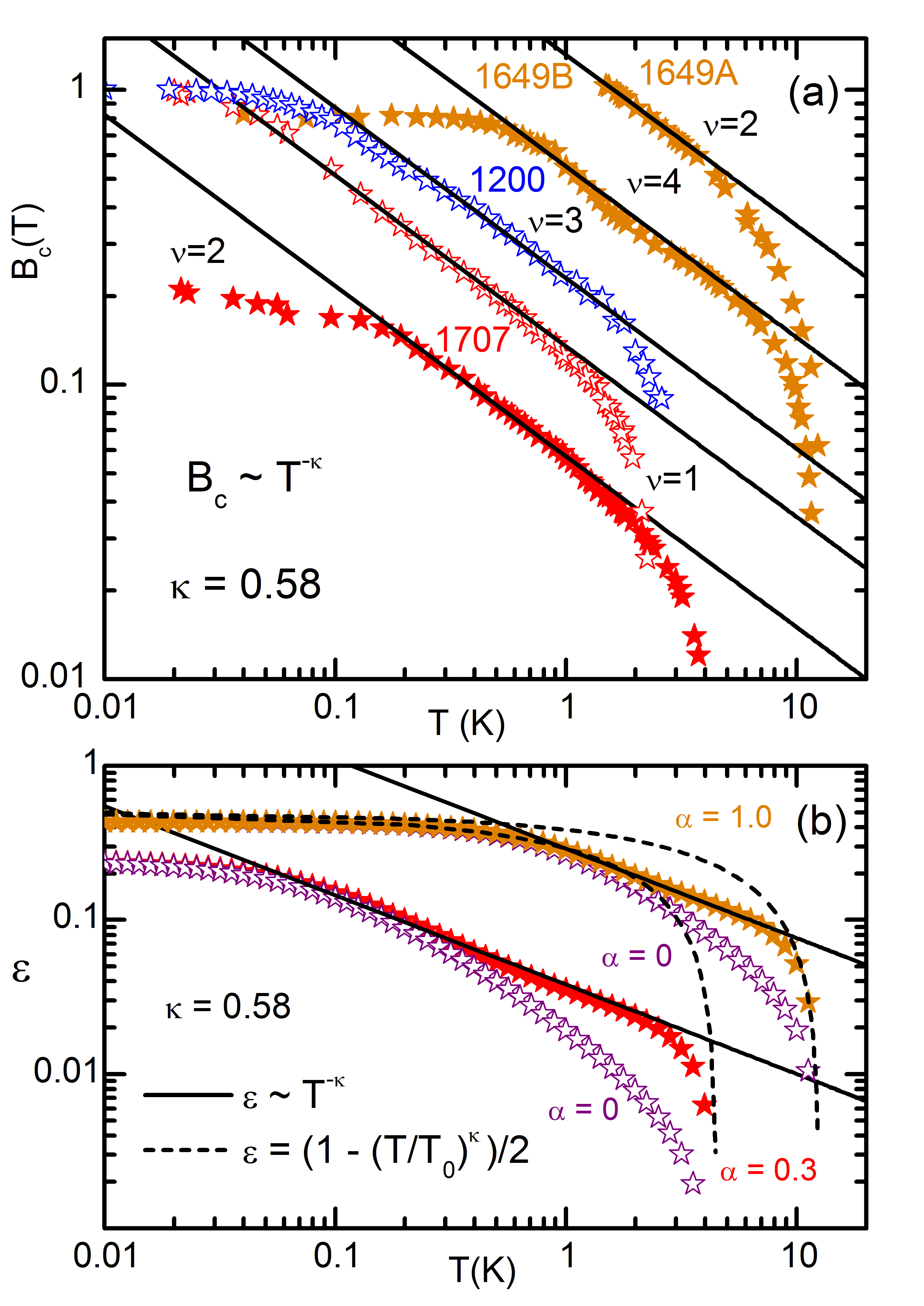}
\end{center}
\caption{(color online) (a) Log log plot of $B_c(T)$ versus temperature for selected odd and even filling factors,
which show a distinct linear regime, for all samples investigated. The solid lines show a power law dependence with
$B_c \propto T^{-\kappa}$ with $\kappa=0.58$. (b) Log log plot of the partial filling factor $\varepsilon$ versus
temperature calculated with the thermal activation model. The parameters used correspond to $\nu=2$ in sample 1707 and
$\nu=4$ in sample 1649B (see Table\,\ref{Tab:FitParam}). The calculations are made with (closed symbols) and without
(open symbols) thermal broadening. The solid lines are a power law behavior $\varepsilon \propto T^{-\kappa}$ with
$\kappa=0.58$. The broken lines are the expected $\varepsilon(T) = (1 - (T/T_0)^\kappa)/2$ from scaling theory as
described in the text.}\label{fig:ScalingAllSamples}
 \end{figure}

Looking at the $B_c(T)$ data for all the samples investigated, it is clear that the lower filling factors have large
regions, often over almost and order of magnitude in magnetic field and temperature, in which the dependence is linear.
This suggests that $B_c(T)$ has a power law dependence. We note, that where the linear behavior persists to higher
filling factors (see \emph{e.g.} inset of figure\,\ref{fig:Even1649Barry}), the slope is always the same. In other
words it appears to be quite universal in the sense it is independent of Landau level index. In order to compare the
$B_c(T)$ data, we plot in figure\,\ref{fig:ScalingAllSamples}(a), using a double log scale, $B_c(T)$ for the lowest odd
and even filling factor which show a linear dependence for the different samples investigated. All the plotted filling
factors, both odd and even have a linear region with identical slopes, suggesting that the power law dependence is
universal, independent of both the Landau level index and the sample. The solid lines are the predicted power law
dependence $B_c \propto T^\kappa$ with $\kappa=0.58$. Clearly, the agreement is remarkable given the identical slope
observed for different filling factors from different samples. There is one exception, while $\nu=3$ in sample 1200 has
a large linear region with a slope close to $\kappa=0.58$, the even filling factors (\emph{e.g.} $\nu=4$ which is not
plotted in figure\,\ref{fig:ScalingAllSamples} for clarity), have a smaller slope $\simeq 0.29$. As our simple intra
Landau level activation model correctly reproduces the data for a number of different samples with very different
carrier densities and disorder, the question arises as to where the apparent universality comes from and what are the
implications for the scaling theory which applies to the width of the maxima in $R_{xx}$?

In order to answer this question, we must first establish a link between $\Delta B$ (width of $R_{xx}$ peak) of scaling
theory and $B_c$ (half width of $R_{xx}$ minimum). In the intra Landau level activation model the problem is symmetric
in filling factor (see section\,\ref{Sec:TherAct}); when moving away from integer filling factor $\nu$, the resistance
ceases to be dissipationless at filling factors $\nu \pm \varepsilon$, corresponding to magnetic fields $B_1=B_f/(\nu +
\varepsilon)$ and $B_2=B_f/(\nu -\varepsilon)$ with $B_c = (B_2 - B_1)/2$. Thus, we can write
\begin{equation}
B_c = \frac{B_f \varepsilon}{\nu^2 (1 - \varepsilon^2/\nu^2)},\nonumber
\end{equation}
which varies to a good approximation as $1/\nu^2$ (in agreement with experiment) with a small correction $(1 -
\varepsilon^2/\nu^2)^{-1} \approx 1$ which deviates appreciably from $1$ only for the lowest filling factor. Thus, $B_c
\simeq B_f \varepsilon/\nu^2$. If we reason in terms of the filling factor, there is no approximation involved with a
width of $2\varepsilon$. The temperature dependence of $B_c$ then arises from the temperature dependence of the partial
filling factor $\varepsilon(T)$. If $B_c(T)$ scales as $T^{-\kappa}$ this implies that $\varepsilon \propto
T^{-\kappa}$ (here we neglect $(1 - \varepsilon^2/\nu^2)^{-1} \approx 1$).

In a similar way the peak in $R_{xx}$ between integer filling factors $\nu$ and $\nu -1$ has a width $\Delta B
=B_f/(\nu + \varepsilon -1) - B_f/(\nu - \varepsilon)$. This can be rearranged to give,
\begin{equation}
\Delta B = \frac{B_f (1 - 2\varepsilon)}{\nu^2 - \nu - \varepsilon(1-\varepsilon)},\nonumber
\end{equation}
which provided $\nu^2 \gg \varepsilon(1-\varepsilon)$ leads to $\Delta B \propto (1-2\varepsilon)$. Again, in terms of
filling factor, there is no approximation involved, with a width of $1 - 2\varepsilon$. If $\Delta B$ is to scale as
$T^\kappa$ this implies that (within the framework of the activation model) $(1 - 2\varepsilon) = (T/T_0)^\kappa$.
Thus, it is mathematically impossible that both $B_c(T)$ and $\Delta B(T)$ simultaneously have a power law dependence;
the observation of a scaling behavior for $B_c$ precludes the observation of a scaling behavior for $\Delta B$ and vice
versa.

In figure\,\ref{fig:ScalingAllSamples}(b) we plot the temperature dependence of $\varepsilon$ calculated using the
thermal activation model with parameters corresponding to $\nu=2$ in sample 1707 and $\nu=4$ in sample 1649B. The
calculations have been made with and without thermal broadening. The solid lines indicate a power law dependence
$\varepsilon \propto T^{-\kappa}$ with $\kappa=0.58$. Without thermal broadening (open symbols), the slope of
$\varepsilon(T)$ (on a log log plot) changes continuously from horizontal at low $T$, to vertical at high $T$. With
such a behavior it is inevitable that, at some point, the slope equals $-0.58$, at least over a limited temperature
range. Including thermal broadening in the calculations (closed symbols) prolongs this behavior to higher temperatures,
creating a wider range of temperatures over which $\varepsilon \propto T^{-\kappa}$ with $\kappa=0.58$. Thus, we
conclude that universality is explicitly absent from the intra Landau level thermal activation model (the slope changes
continuously as a function of temperature), and the observed power law behavior is generated by a sample (but not
filling factor) dependent thermal broadening parameter. A power law dependence of $\Delta B$ would require  $(1 -
2\varepsilon) = (T/T_0)^\kappa$ which can be rearranged to give $\varepsilon = (1 - (T/T_0)^\kappa)/2$. The broken
lines in figure\,\ref{fig:ScalingAllSamples}(b) show the required behavior of $\varepsilon$. Scaling theory implicitly
assumes that all states are localized at $T=0$ ($\varepsilon=0.5$), which is not what is found experimentally. However,
the functional form is roughly correct; flat at low temperature and falling off rapidly at high temperature. Note that
here, the $T_0$ parameter of scaling theory plays the role of the critical temperature at which the dissipationless
conductance ceases to exist.

\section{Conclusion}\label{Sec:Conclusion}

A simple model involving thermal activation, from localized states in the tail of the Landau level at the Fermi energy,
to delocalized states above the mobility edge in the same Landau level, explains the $B_c(T)$ phase diagram for a
number of different quantum Hall samples with widely ranging carrier density, mobility and disorder. Good agreement is
achieved over $2-3$ orders of magnitude in temperature and magnetic field for a wide range of filling factors. The
width of the low temperature $B_c(T)$ region depends sensitively on the Landau level broadening $\Gamma$. For a given
sample, both even and odd filling factors can be fitted with the same value of $\Gamma$ demonstrating that the Landau
level width is independent of magnetic field in the high field regime. The position of the mobility edge is also
independent of magnetic field, provided the Landau level overlap does not change. Our data suggest that the mobility
edge moves to maintain a sample dependent critical density of states at that energy, leading to a simple relation
between the position of the mobility edge with and without the opening of the spin gap. The same model can also be
applied to fractional quantum Hall states via the composite Fermion model. The composite Fermion Landau levels have
exactly the same width as their electron counterparts, as previously suggested based upon a Dingle analysis of the low
field composite Fermion oscillations.\cite{Leadley1994} The long tails of a Lorentzian are essential for the activation
model, providing localized states deep in the gap, which are required to reproduce the robust high temperature part of
the $B_c(T)$ phase diagram. This is in agreement with published torque measurements, the detailed analysis of which
concluded that Lorentzian broadening provided the best fits to the saw tooth like oscillations in the 2DEG
magnetization.\cite{Potts1996}

\begin{acknowledgments}
This work was partially supported by ANR JCJC project milliPICS, the Region Midi-Pyr\'en\'ees under contract MESR
13053031 and IDEX grant BLAPHENE.
\end{acknowledgments}


%

\end{document}